\documentclass[sigconf,anonymous=false]{acmart}

\usepackage{booktabs} % For formal tables
\usepackage{lipsum} 
\usepackage{balance}
\usepackage{paralist}
\usepackage{filecontents}
\usepackage{etoolbox}
\usepackage{mathrsfs}
\usepackage{bm}
\usepackage{setspace}
\usepackage{amssymb}
\usepackage{graphicx}
\usepackage{caption}
\usepackage{subcaption}
\usepackage{multirow}
\setcopyright{rightsretained}
\usepackage[utf8]{inputenc}
\usepackage[english]{babel}

\newtheorem{theorem}{Condition}

\begin{document}
\title{A Zero Attention Model for Personalized Product Search}

\copyrightyear{2019} 
\acmYear{2019} 
\acmConference[CIKM '19]{The 28th ACM International Conference on Information and Knowledge Management}{November 3--7, 2019}{Beijing, China}
\acmBooktitle{The 28th ACM International Conference on Information and Knowledge Management (CIKM '19), November 3--7, 2019, Beijing, China}\acmDOI{10.1145/3357384.3357980}
\acmISBN{978-1-4503-6976-3/19/11}

\settopmatter{printacmref=true, printfolios=false}
\fancyhead{}
%\titlenote{Produces the permission block, and
%  copyright information}
%\subtitle{Extended Abstract}
%\subtitlenote{The full version of the author's guide is available as
%  \texttt{acmart.pdf} document}

\iftrue

\author{Qingyao Ai}
\authornote{This work is done in Amazon Search during the first author's Ph.D. at UMass Amherst.}
\affiliation{%
	\institution{School of Computing, University of Utah}
	%\streetaddress{P.O. Box 1212}
	\city{Salt Lake City} 
	\state{UT} 
	\country{USA}
	%\postcode{01003-9264}
}
\email{aiqy@cs.utah.edu}
\author{Daniel N. Hill}
%\authornote{The secretary disavows any knowledge of this author's actions.}
\affiliation{%
	\institution{Amazon Search}
	%\streetaddress{P.O. Box 1212}
	\city{Berkeley} 
	\state{CA} 
	\country{USA}
	%\postcode{01003-9264}
}
\email{daniehil@amazon.com}
\author{S. V. N. Vishwanathan}
%\authornote{The secretary disavows any knowledge of this author's actions.}
\affiliation{%
	\institution{Amazon Search}
	%\streetaddress{P.O. Box 1212}
	\city{Palo Alto} 
	\state{CA} 
	\country{USA}
	%\postcode{01003-9264}
}
\email{vishy@a9.com}

\author{W. Bruce Croft}
%\authornote{The secretary disavows any knowledge of this author's actions.}
\affiliation{%
	\institution{CICS, UMass Amherst}
	%\streetaddress{P.O. Box 1212}
	\city{Amherst} 
	\state{MA} 
	\country{USA}
	%\postcode{01003-9264}
}
\email{croft@cs.umass.edu}
\fi

\begin{abstract}

Product search is one of the most popular methods for people to discover and purchase products on e-commerce websites.
Because personal preferences often have an important influence on the purchase decision of each customer, it is intuitive that personalization should be beneficial for product search engines.
While synthetic experiments from previous studies show that purchase histories are useful for identifying the individual intent of each product search session, the effect of personalization on product search in practice, however, remains mostly unknown.
In this paper, we formulate the problem of personalized product search and conduct large-scale experiments with search logs sampled from a commercial e-commerce search engine.
Results from our preliminary analysis show that the potential of personalization depends on query characteristics, interactions between queries, and user purchase histories.
Based on these observations, we propose a Zero Attention Model for product search that automatically determines when and how to personalize a user-query pair via a novel attention mechanism.
Empirical results on commercial product search logs show that the proposed model not only significantly outperforms state-of-the-art personalized product retrieval models, but also provides important information on the potential of personalization in each product search session.

\end{abstract}

\iffalse

\fi
% We no longer use \terms command


\maketitle

%!TEX root=CIKM19-ZAM.tex
\section{Introduction}

%Product search is important for e-shopping

%purchase is personal, so personalization is important for product search

%Is this really true? example. in web search, this is not true. May be it is same in product search

%the problem of previous studies on personalization in product search always to personalization equally, e.g. Ai et al.

%In this paper, we want to explore the effect of personalization in product search and learn when and how to choose personalization weights.
%to do so, first log analysis, results
%then model, idea, results

%contribution

%paper structure

%E-shopping is an important part of people's lives.
%Due to the overwhelming number of products online, search and recommendation are the major methods to discover products on e-commerce websites.
%According to a recent survey\footnote{https://www.bigcommerce.com/blog/ecommerce-trends/#top-19-ecommerce-trends-of-2018}, the actual traffic of recommender systems on  

Due to the increasing popularity of online shopping and a large number of products on e-commerce websites, product search has become one of the most popular methods for customers to discover products online.
%Product search is one of the most popular methods for people to find and shop products online. 
In a typical product search scenario, a user would first issue a query on the e-commerce website to get a list of relevant products, then browse the result page, and select one or more items to purchase. 
Therefore, the quality of product search results has a direct impact on customer satisfaction and the number of transactions on e-commerce websites.

Because purchasing is a personal behavior with a real financial cost, it is well-recognized that personal preferences could directly affect customer's purchase decisions~\cite{moe2003buying}.
Previous studies show that many purchase intents in product search can be influenced by a user's personal taste and experience~\cite{Sondhi:2018:TQE:3209978.3210152,su2018user}.
Experiments on synthetic data also demonstrate that incorporating user information extracted from product reviews and purchase histories can significantly improve the performance of product retrieval models~\cite{ai2017learning}.
Thus, it is intuitive that personalization should have a significant potential in product search.

%Because users are choosing products based on their personal needs, it is intuitive to believe that personalization is important for product search experiences.
%In fact, previous studies~\cite{ai2017learning} have shown that incorporating user information into a product retrieval model can significantly improve the performance of a product search system.

%e-shopping is important
%product search and recommendation is the key way to 
%according to report, rec only 30%
%so product is dominate and important
%product search is a scenario xxx
%because users have different preferences, personalization is useful
%previous studies 

A major question is whether personalization always improves the quality of product search.
Attractive as it seems to be, search personalization has been shown to potentially have negative effects by previous studies on Web search~\cite{teevan2008personalize,bennett2012modeling, dumais2016personalized}.
When a customer submits the query ``toothpaste'' to an e-commerce search engine, it is possible that they want personalized search results which suit their personal needs (e.g., sensitive teeth). It is also possible that they do not differentiate much between toothpastes, and simply purchase the best seller on the list.
%for users with non-personal search intents could damage the final performance of the search engine.
On the one hand, personalization with user information can help us better understand the user's search intent when there is limited information revealed in the query.
On the other hand, incorporating unreliable personal information could exacerbate the problem of data sparsity and introduce unnecessary noise into a search model. 
When and how to conduct personalization is an important research question for product search in practice.

Despite its importance, the effect of personalization on real product search engines has not been extensively studied.
To the best of our knowledge, existing work on personalized product search only uses user information as an additional feature in the retrieval model, and conducts undifferentiated personalization in all search sessions~\cite{jannach2017investigating,ai2017learning}.
This research constructs user profiles with query-independent information and provides little insight on how personalization would benefit or damage the performance of product retrieval in different search scenarios.  
%Also, to the best of our knowledge, most existing personalization techniques for product search construct user profile in  

In this paper, we explore the potential and risks of personalization for product search.
Specifically, we focus on the question of when and how to do search personalization with a user's purchase history on e-commerce websites.
We start from a theoretical analysis and identify two necessary conditions in order for personalization to benefit product search.
Then, we verify the two conditions on large-scale search logs sampled from a commercial e-commerce search engine.
We find that, while personalization appears to be useful for queries with medium or high frequency, it tends to be less beneficial on tail queries with low frequency.
Also, we notice that the importance of personalization in product search often depends on the interactions between query context and the user's previous purchases.
It is impossible to determine the usefulness of personalization without knowing the query and the user's purchase history simultaneously.
%Despite its potentials, personalization has not been extensively studied in product search so far.
%The effect of search personalization on real product search traffic remains mostly unknown.

Based on our observations, we propose a Zero Attention Model for differentiated personalization in product search. 
%Different from previous studies~\cite{ai2017learning} where user preferences and query intents are modeled independently, our Zero Attention Model conduct query-dependent depends on both 
Previous studies on personalized product search often model users and queries separately by assuming that they are independent~\cite{ai2017learning}. 
%This, however, is not true because users' preferences on products could vary according to their search intent in the current session.   
%This, however, is problematic when there are many items in a user's purchase history that are not useful for the personalization of the current query. 
%This, however, makes it impossible to capture the users' preferences specifically related to their current search intent.
This, however, makes it impossible to conduct query-specific personalization in each search session.
To that end, we propose a Zero Attention Strategy that constructs user profiles as a weighted combination of their previously purchased items. %based on how relevant each item is with respect to the current search query.
The idea of the Zero Attention Strategy is to allocate different attention to the user's previous purchases according to their current search intent.  
More importantly, in contrast to a classic attention mechanism where the model must attend to at least one item in the input, the Zero Attention Strategy introduces a zero vector and allows the attention model to pay no attention to any input.
As shown in Section~\ref{sec:attention_model}, this essentially creates a threshold function that enables our product retrieval model -- the Zero Attention Model (ZAM) -- to automatically determine when and how to personalize search results based on the current query and user information.
Our experiments on real e-commerce search logs demonstrate that ZAM can significantly outperform state-of-the-art baselines for personalized product search.
Also, the attention weight on the zero vector in ZAM is a good indicator for the potential of personalization in each search session.
%Experiments on real search logs not only demonstrate the superior effectiveness of ZAM comparing to the state-of-the-art personalized product retrieval models, but also show that the attention weight on the zero vector is a good indicator for the potential of personalization for each user-query pair.

\iftrue
In summary, the major contributions of this paper include:
\begin{itemize}
\item We present both a theoretical and empirical analysis of the potential of personalization in product search with large-scale search logs sampled from a commercial e-commerce search engine.
\item We propose a novel Zero Attention Model for personalized product search by conducting differentiated personalization for different query-user pairs.
\item We conduct comparisons over state-of-the-art personalization techniques for product search, and discuss their advantages and disadvantages in practice. 
\end{itemize}
\fi

\section{Related Work}\label{sec:related_work}

Our work in this paper is closely related to the research of product search, search personalization, and neural retrieval models.

\textbf{Product Search}. 
Product search is an important problem that has been widely studied in the research communities of Data Mining and Information Retrieval.
Early studies mainly focus on how to construct effective retrieval systems that support search for structured product information~\cite{duan2013supporting,duan2013probabilistic,duan2015mining}. 
For example, Lim et al.~\cite{lim2010multi} propose to conduct product retrieval with a facet search engine built on structured representations of products in relational databases such as brands, prices, categories, etc. 
%In practice, however, e-commerce website users tend not to search with structured queries in most cases because it requires considerable domain knowledge is time-consuming~\cite{duan2013supporting}.
%Instead, the most popular paradigm for modern product search engine is to directly retrieve products based on natural language queries.
%Duan et al.~\cite{duan2013supporting,duan2013probabilistic,duan2015mining} propose a mixture model with language modeling approaches to enrich query and product representations so that we can conduct conditional search on product specifications.
Later, product search studies move to more advanced retrieval models that support more complicated search data and optimization objectives.
For instance, Van Gysel et al.~\cite{van2016learning} introduce a latent semantic entity model that matches products and queries in a latent semantic space; Guo et al.~\cite{guo2018multi} propose a TranSearch model that can search product images with text queries.
There are also a variety of studies on applying feature extraction and learning-to-rank techniques~\cite{wu2017ensemble,aryafar2017ensemble,karmaker2017application,Hu:2018:RLR:3219819.3219846} to product search for the optimization of different product retrieval metrics~\cite{wu2018turning}. 
In this paper, we focus on the problem of when and how to conduct search personalization in product retrieval. 
%However, because of the language gaps between how customers formulate queries and how sellers write product descriptions~\cite{nurmi2008product},  

%product search (no personalization)
%duan's work, KDD, SIGIR 

\textbf{Search Personalization}.
Studies on search personalization focus on providing different results to each individual to maximize their search experience~\cite{dumais2016personalized}.
In the context of Web search, this means re-ranking documents according to the personal needs of each user based on their locations, search histories, clicked documents, etc.~\cite{chirita2005using,shen2005implicit,teevan2005personalizing,bennett2012modeling}.
While search users could have different behavior for the same query~\cite{dou2007large,white2007investigating}, it has been found that personalization is not always beneficial for the effectiveness of search engines.
For example, in a large-scale log analysis of user behavior patterns on Live Search, Teevan et al.~\cite{teevan2008personalize} identify multiple factors that affect the usefulness of personalization in Web search, such as result entropy, result quality, search tasks, etc.
%To evaluate the potentials and risks of personalization in Web search, 
%It is useful according to user behaivor  
%personalization, first web search, then product search

In the scope of product search, however, the effect of search personalization has not been fully studied. 
%Because purchase is a personal behavior with large cost to each individual, it is commonly believed that personal preferences could significantly affect the purchase decision of search users~\cite{ai2017learning}.
While recent studies on e-commerce search logs~\cite{Sondhi:2018:TQE:3209978.3210152,su2018user} identify and analyze several types of product search intents qualitatively, there has been little work on quantifying the potentials and risks of personalization 
in product search.
There are two notable exceptions.
The first is Jannach and Ludewi's work on applying personalized recommendation techniques to product search~\cite{jannach2017investigating}. The authors reduce the problem of product search to a recommendation task by ignoring the search query in the ranking. A second study by Ai et al. investigated personalized product search using Amazon Review datasets ~\cite{ai2017learning}. They conducted experiments with synthetic queries generated from product category information. To the best of our knowledge, our work is the first study that analyzes the impact of personalization on a real e-commerce search engine. 

\textbf{Neural Retrieval Models}.
The advance of deep learning techniques has attracted much attention in the IR community recently.
Neural retrieval models are highly effective as they can automatically learn and incorporate embedding features in ranking optimizations~\cite{huang2013learning,shen2014latent,guo2016deep,mitra2017learning,zamani2017situational}.
%A variety of neural retrieval models have been proposed for IR, such as the representation-based models~\cite{huang2013learning,shen2014latent} and interaction-based models~\cite{guo2016deep,mitra2017learning}.
For product search, Van Gysel et al.~\cite{van2016learning} propose a latent semantic entity model that jointly learns the embeddings of words, queries, and items so that products could be directly retrieved according to their similarities in the latent space.
Later, Ai et al.~\cite{ai2017learning} incorporate user information extracted from user purchases and product reviews into a neural generative framework and create a Hierarchical Embedding Model for product search. 

Attention is an important deep learning technique that enables neural networks to allocate variable attention or weighting to different model components in the training or prediction process~\cite{vaswani2017attention}.
It has proved effective in many AI applications such as Computer Vision~\cite{mnih2014recurrent,chen2015abc}, Natural Language Processing~\cite{rush2015neural,luong2015effective}, etc.
Examples in e-commerce studies include the work of
Chen et al.~\cite{chen2018sequential} which applies attention mechanisms to the construction of a personalized recommender system. 
Their task, however, is fundamentally different from product search as users cannot specify their information needs explicitly in recommendation scenarios.  
In this paper, we focus on the problem of personalized product search and propose a new attention mechanism that allows a retrieval model to conduct differentiated personalization based on the previous purchases of the user and their relationship to the current query.

% deep learning is popular
% neural retrieval models have been proposed 
% For product search, neural retrieval models
% Attention mechanism in popular
% attention mechanism in recommendation
% In this paper, we propose new for personalized product search.

%neural retrieval models, embedding models for product search, attention

%!TEX root=CIKM19-ZAM.tex
\section{Preliminary Analysis}\label{sec:log_analysis}

\iftrue
\begin{figure*}
	\centering
	\begin{subfigure}{.45\textwidth}
		\centering
		\includegraphics[width=2.5in]{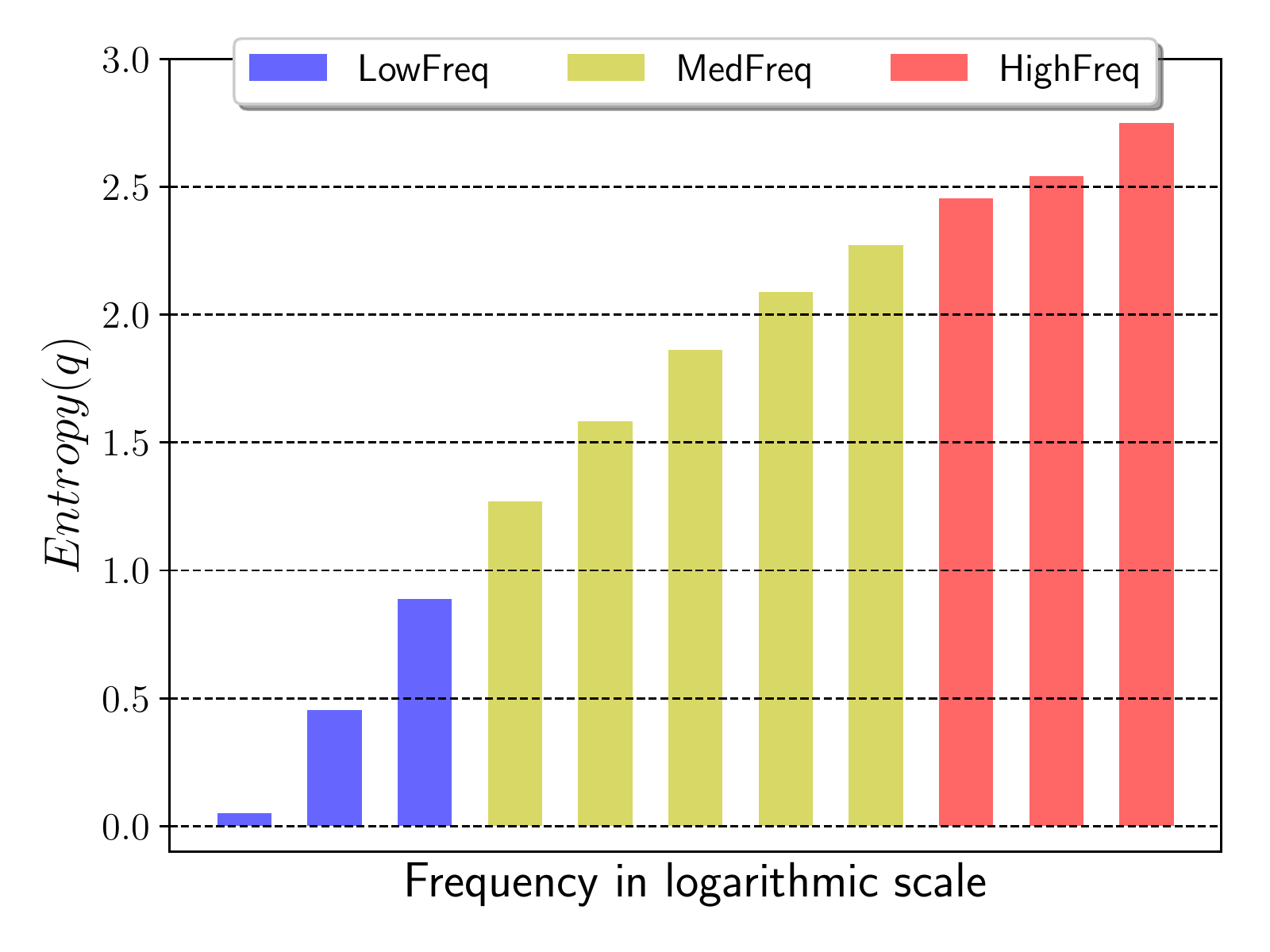}
		\caption{Purchase Entropy}
		\label{fig:entropy}
	\end{subfigure}
	\begin{subfigure}{.45\textwidth}
		\centering
		\includegraphics[width=2.5in]{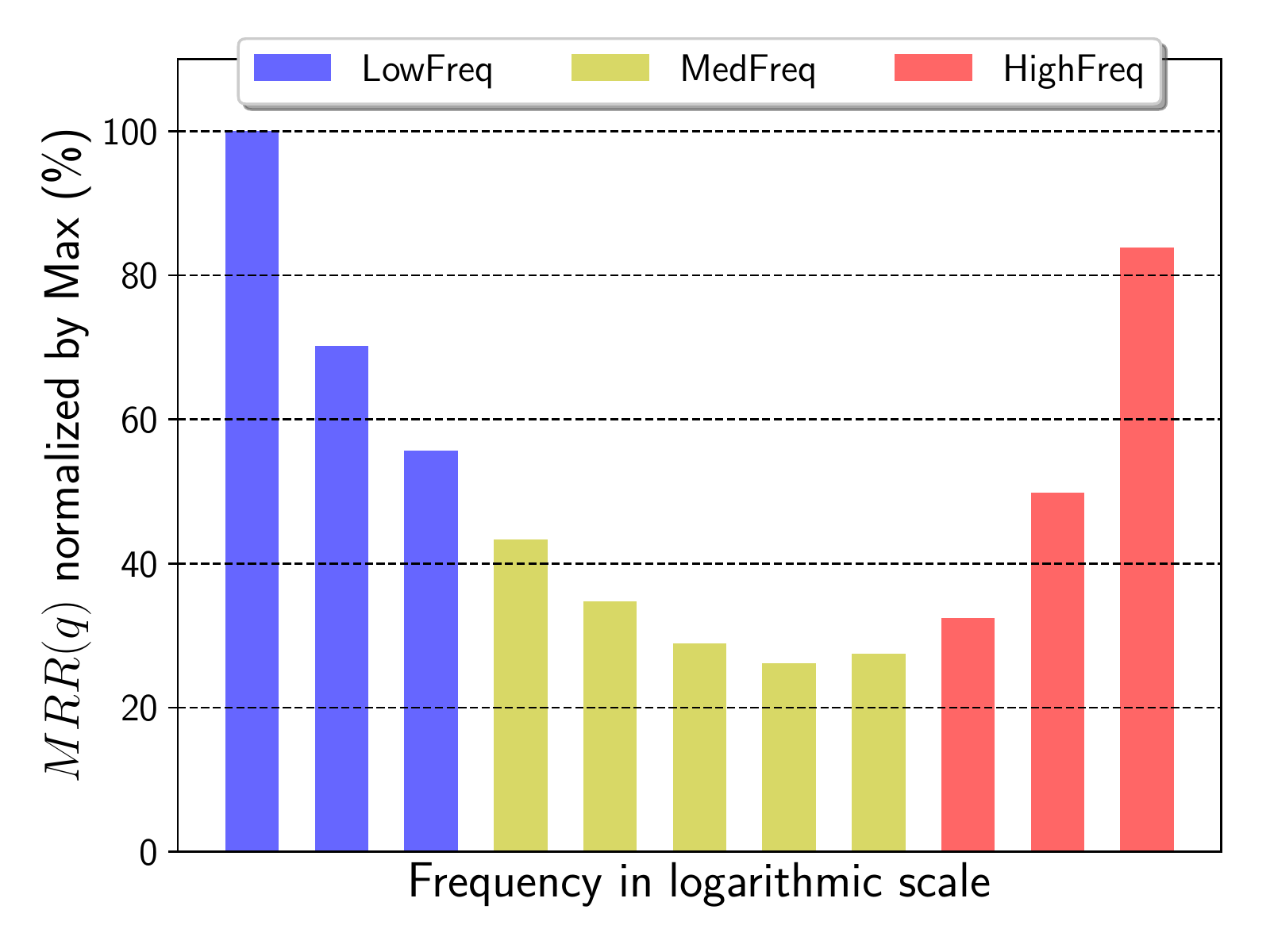}
		\caption{Popularity Model Performance}
		\label{fig:frequency_perfermance}
	\end{subfigure}%
	\caption{The purchase entropy $Entropy(q)$ and popularity model performance $MRR(q)$ for queries with different frequencies.}
	\vspace{-10pt}
\end{figure*}

\fi
\iffalse
\begin{figure}
	\centering
	\includegraphics[width=2.5in]{./figure/entropy.pdf}
	\caption{The entropy of queries with different frequencies. }
	\label{fig:entropy}
\end{figure}
\fi

% intro
In this section, we present our preliminary analysis on the potential of personalization in product search.
Previous work has demonstrated that incorporating user information is beneficial for the performance of product retrieval models on synthetic data generated from product category hierarchies~\cite{ai2017learning}. However, to the best of our knowledge, there is no systematic study on why and how personalization could help product search in practice. 
%Ai et al. have shown that incorporating user information is beneficial for the performance of product retrieval models.
%They, however, focused on a synthetic dataset generated from product category information.
To fill this gap, we conduct a theoretical analysis of the fundamental assumptions behind personalized product search, and validate them on the search logs sampled from a commercial e-commerce search engine.
More details about the log data can be found in Section~\ref{sec:setup}.
%Specifically, we randomly sampled one million users from the product categories of \textit{Beauty}, \textit{Grocery}, and \textit{Health\&Personal Care}, and gathered both their queries and corresponding purchases from Septermber 2017 to July 2018.
%To fill this gap, we conduct a log analysis with the search logs of one million customers sampled from a commercial e-commerce search engine from Septermber 2017 to July 2018. 
%We develop and test two hypotheses on the log data to evaluate the potential of personalization in different search scenarios.

The key difference between personalized and non-personalized product search is the modeling of the user's purchase behavior. 
Formally, let $q$ be a query submitted by a user $u$.
In a personalized product retrieval model, purchase behavior on an item $i$ depends on both the current search context and the user's personal preferences. 
The probability of whether $i$ would be purchased by $u$ in the query $q$ should be parameterized as $P(i|q,u)$. 
By contrast, non-personalized product retrieval models assume that user's purchase behavior only depends on the current search context (i.e., query $q$).
It models the probability of item purchase as $P(i|q)$, which can be treated as the aggregated group preference over all possible users as
$$
P(i|q) = \sum_{u \in \mathcal{U}} P(i|q,u)\cdot P(u)
$$
where $\mathcal{U}$ is the universal user set, and $P(u)$ is the probability of observing a specific user $u$ in the search logs.
We assume that the distribution of $u$ is independent of $q$ for simplicity.

Given this formulation, we can identify the first necessary condition for personalization to benefit product search as:
\begin{theorem}
%Personalization is benefitical only when there are more than one type of products relevant to the query.
Personalization is beneficial only when the query carries diverse purchase intents.
%With same query, different users would purchase different items.
\end{theorem}
%\noindent In other words, in order for $P(i|q,u)$ to be a better model than $P(i|q)$, more than one item have to be purchased by users who search with $q$. 
%The proof is straightforward -- there is no need to do personalization when all users purchase the same product in a particular query. 
\noindent The proof is straightforward -- $P(i|q,u)$ cannot be a better model than $P(i|q)$ when $q$ only represents the intent to purchase a specific item. 
The more specific intent a query has, the less beneficial personalization could be in product search.

To analyze query specificity, we compute the purchase entropy of each query in the sampled e-commerce search logs as 
\begin{equation}
Entropy(q) = -\!\!\sum_{i \in I_q}\!\!P(i|q)\log P(i|q)=  -\!\!\sum_{i \in I_q}\!\!\frac{\#(i,q)}{|S_q|}\log_2(\frac{\#(i,q)}{|S_q|})
\label{equ:entropy}
\end{equation}
where $I_q$ is the candidate item set for query $q$, $S_q$ represents the search sessions for $q$, and $\#(i,q)$ refers to the number of sessions in $S_q$ where item $i$ has been purchased.

Figure~\ref{fig:entropy} shows the purchase entropy of queries on \textit{Beauty} products (e.g., facial cleanser) in our sampled search logs. %on queries with different frequency in  \textit{Beauty}. 
%Since we cannot reveal the actual volume of search traffic due to the privacy policy, 
Here, we rank queries according to their frequencies in logarithmic scale, and split them into three groups: the queries with low frequency (\textit{LowFreq}), with medium frequency (\textit{MedFreq}), and with high frequency (\textit{HighFreq}). 
The group sizes are balanced so that the total number of sessions in each group is approximately the same.
As depicted in the figure, purchase entropy shows a strong correlation with query frequency. 
When the number of sessions increases, we are likely to observe more purchases on different items, which usually indicates diverse purchase intents.
Hence, it seems that queries with high frequencies have more potential for personalization.
%It seems intuitive to believe that queries with high frequencies tend to be less specific and have more potential for personalization.

This, however, may not be true.
While a high purchase entropy indicates a high diversity of purchase intents, it doesn't necessarily mean that every user has unique preferences in search.
Therefore, the second necessary condition for the effectiveness of personalization in product search is:
\begin{theorem}
Personalization is beneficial only when the personal preferences of individuals are significantly different from their aggregated group preference. 
%The personal preferences of individual users should be significantly different from the aggregated group preferences.
\end{theorem}
\noindent In other words, personalized models can outperform non-personalized models only when the distribution of $P(i|q,u)$ is significantly different from $P(i|q)$.

A simple method to evaluate the differences between $P(i|q,u)$ and $P(i|q)$ in search problems is to rank items with $P(i|q)$ and evaluate the performance with $P(i|q,u)$. 
Specifically, we use:
\begin{equation}
MRR(q) = \sum_{u \in \mathcal{U}}\!\!RR(P(i|q), P(i|q,u)) \cdot P(u) 
\label{equ:frequency_performance}
\end{equation} 
where $RR(P(i|q), P(i|q,u))$ is the reciprocal rank of a ranked list produced by ranking with $P(i|q)$, using $P(i|q,u)$ as the ground truth. 
In other words, $MRR(q)$ can be computed as the mean reciprocal rank of a retrieval model that ranks products according to how many times they have been purchased given the query (i.e., \textit{the Popularity Model}), using the actual user purchases in each session as the ground truth.
%Although $MRR(q)$ is not strictly correlated with $H(q)$, it is numerically stable and can directly reflect the similarity between $P(i|q,u)$ and $P(i|q)$ in ranking.
%the closer the distribution of $P(i|q)$ is to $P(i|q,u)$, the larger $MRR(q)$ will be.
The higher $MRR(q)$ is, the closer $P(i|q,u)$ is to $P(i|q)$ and the less beneficial search personalization can be.

Figure~\ref{fig:frequency_perfermance} shows the $MRR(q)$ of queries on \textit{Beauty} products.
We compute the Popularity Model based on one year of search logs and test it on the search data sampled from the following week. 
Numbers are normalized with respect to the maximum $MRR(q)$ of all query groups.   
%We report the relative improvements of $MRR(q)$ in each query group with respect to the global MRR of the standard language modeling approach (QL)~\cite{ponte1998language} on all queries (more details can be found in Section~\ref{sec:setup}).
As we can see in the figure, $MRR(q)$ shows a U shape with respect to query frequency. 
While tail queries and head queries tend to have high $MRR(q)$, queries with median frequencies usually have low $MRR(q)$ in our search logs.
This indicates that, unlike purchase entropy, the similarity between $P(i|q,u)$ and $P(i|q)$ is not monotonically correlated with query frequency.

To further understand the results of Figure~\ref{fig:entropy} and Figure~\ref{fig:frequency_perfermance} together, we extract some example queries from the search logs for each query group. %of queries in Table~\ref{tab:query_examples}.
In \textit{LowFreq}, we observe many \textit{spear-fishing} queries where the user directly specifies the name of the item in the query string, such as ``stila stay all day waterproof liquid eyeliner'', ``mario badescu drying lotion'', etc.
When searching with these queries, users usually have exact items in mind.
Thus, the query specificity is high and ranking by $P(i|q)$ achieves a good performance in general. 
In \textit{MedFreq}, we observe queries on personal products like ``body wash for women'', ``maybelline mascara'', etc.
Although the concept of relevance could be personal in these queries, users only provide vague descriptions of what items they want in the query strings.
%As a result, the purchase entropy is high and the similarity between $P(i|q,u)$ and $P(i|q)$ in ranking is low.
As a result, the purchase entropy is high and the $MRR(q)$ of the Popularity Model is low. 
In \textit{HighFreq}, while we observe many queries on popular products in daily activities, such as ``sunscreen'', ``nail file'', and ``mouthwash'', the user behavior patterns in these queries, however, are not consistent.
For ``sunscreen'', we observe products with a variety of properties such as ``oil-free'', ``water-resistant'', etc. 
It seems easy to identify the need of each user using the products they purchased.
Whereas, in ``nail file'' and ``mouthwash'', most candidate items are cheap and homogeneous.
While the purchase entropy of these queries is high, we haven't identify any salient personal preferences over the candidate products among the customers we sampled. 
In many cases, the customers simply purchased one of the best sellers.
Therefore, the potential of personalization varies significantly in different queries, and this motivates us to develop a more sophisticated model for personalization in product search.
%Therefore, the potential of personalization are not perfectly correlated with simple query characteristics such as frequency and specificity, and this motivates us to develop a more sophisticated model for personalization in product search.
%Therefore, the potentials of personalization in product search vary significantly in different sessions depending on both the query characteristics and the strength of user preferences in the search contexts.

%To further understand the results of Figure~\ref{fig:entropy} and Figure~\ref{fig:frequency_perfermance} together, we conduct case studies on search logs and show the example queries in each group of queries in Table~\ref{tab:query_examples}.
%In \textit{LowFreq}, we observe many spear-fishing queries that directly specify the name of an item in the query strings, such as xxx, xxx, etc.
%On these queries, users usually have exact items in mind.
%Thus, the purchase entropy is low, and ranking by $P(i|q)$ could achieve a good performance in general. 
%In \textit{HighFreq}, we observe a lot of queries on popular products in daily activities such as toothpaste, xxx, etc.
%While the purchase entropy is high, most candidate items in these queries are cheap and homogeneous.
%It is difficult to find any salient personal preferences on users who submitted these queries.
%In \textit{MedFreq}, however, we observe a variety of queries on personal products like xxx, xxx, and xxx.

% Formulation

% Assumption 1

% Assumption 2

% Case study

% conclusion and next

%!TEX root=CIKM19-ZAM.tex
\section{Zero Attention Model}\label{sec:model}

In this section, we propose a Zero Attention Model (ZAM) for personalized product search.
%The goal of ZAM is to automatically determine when and how to do personalization in product search. 
%ZAM is designed under an embedding-based generative framework, and it conducts query-dependent personalization by constructing user profiles with an attention mechanism.
%Specifically, we propose a Zero Attention Strategy that enables ZAM to automatically decide when and how to attend in different search scenarios.
ZAM is designed under an embedding-based generative framework. 
It conducts query-dependent personalization by constructing user profiles with a Zero Attention Strategy that enables it to automatically decide when and how to attend in different search scenarios.
Theoretical analysis shows that our proposed attention strategy can create a dynamic threshold that controls the weights of personalization based on both the query and the user purchase history.

\subsection{Embedding-based Generative Framework}\label{sec:framework}

Latent semantic models have proven effective for product search and recommendation~\cite{van2016learning,he2017neural}.
Among different types of latent semantic models, neural embedding models have achieved the state-of-the-art performance on many benchmark product search datasets~\cite{ai2017learning,guo2018multi}.
Specifically, Ai et al.~\cite{ai2017learning} propose an embedding-based generative framework that can jointly learn the embedding representations of queries, users, and items by maximizing the likelihood of observed user purchases.

Let $q$ be a query submitted by a user $u$, $i$ be an item in the candidate set $I_q$ for $q$, and $\alpha$ be the size of an embedding vector.
Under an embedding-based generative framework~\cite{mikolov2013distributed,le2014distributed}, the probability of whether $i$ would be purchased by $u$ given $q$ can be modeled as
\begin{equation}
P(i|u,q) = \frac{\exp (\bm{i} \cdot \bm{M_{uq}})}{\sum_{i'\in I_q}\exp (\bm{i}' \cdot \bm{M_{uq}})}
\label{equ:P_iuq}
\end{equation} 
where $\bm{i}\in \mathbb{R}^{\alpha}$ is the embedding representation of $i$, and $M_{uq}$ is a joint model of user-query pair $(u,q)$.
Products are ranked according to $P(i|u,q)$ so that the probability of user purchases in each search session can be maximized.
Depending on the definition of $M_{uq}$, we could have multiple types of embedding-based retrieval models for product search.
Here, we introduce two representative models from previous studies -- the Query Embedding Model and the Hierarchical Embedding Model.

%Products are ranked according to $P(i|q)$ so the probability of user purchases under a specific query can be maximized.
%Here, we introduce two representative neural embedding models for product search: the Query Embedding Model and the Hierarchical Embedding Model.

%QEM
\textbf{Query Embedding Model}. 
The Query Embedding Model (QEM) is an embedding-based generative model for non-personalized product search~\cite{ai2017learning}.
It defines $M_{uq}$ as 
\begin{equation}
\bm{M_{uq}} = \bm{q}
\label{equ:QEM_Muq}
\end{equation}
where $\bm{q}\in \mathbb{R}^{\alpha}$ is the embedding representation of the query $q$. 

Because queries usually are unknown beforehand, $\bm{q}$ must be computed in product search at request time.
Previous studies~\cite{van2016learning,ai2017learning} have explored several methods to construct query embeddings from query words directly. 
One of the state-of-the-art paradigms is to compute query embeddings by encoding query words with a non-linear projection function $\phi$ defined as
\begin{equation}
\bm{q} =  \phi(\{w_q | w_q \in q\}) = \tanh(\bm{W}_{\phi} \cdot \frac{\sum_{w_q \in q}\bm{w_q}}{|q|} + \bm{b}_{\phi})
\label{equ:fs}
\end{equation}
where $\bm{w_q}\in \mathbb{R}^{\alpha}$ is the embedding of a word $w_q$ in $q$, $|q|$ is the length of the query, and $\bm{W}_{\phi}\in\mathbb{R}^{\alpha \times \alpha}$ and $\bm{b}_{\phi}\in\mathbb{R}^{\alpha}$ are two parameters learned in the training process.

%Let $q$ be a query, $i$ be an item in the product candidate set $I_q$ for query $q$, and $\alpha$ be the size of an embedding vector.
%In QEM, the probability of whether item $i$ would be purchased under a query $q$ is defined as
%\begin{equation}
%P(i|q) = \frac{\exp (\bm{i} \cdot \bm{q})}{\sum_{i'\in I_q}\exp (\bm{i}' \cdot \bm{q})}
%\label{equ:QEM_goal}	
%\end{equation}
%where $\bm{i}\in \mathbb{R}^{\alpha}$ and $\bm{q} \in \mathbb{R}^{\alpha}$ are the embedding representations of $i$ and $q$, respectively. 
%Products are ranked according to $P(i|q)$ so the probability of user purchases under a specific query can be maximized.

In QEM, item embedding is learned from their associated text data.
%For example, Ai et al.~\cite{ai2017learning} proposes to learn the embedding of items based on the words in their descriptions.
Let $T_i$ be a set of words associated with an item $i$ (e.g., titles). 
Ai et al.~\cite{ai2017learning} propose to learn $\bm{i}$ by optimizing the likelihood of observing  $T_i$ given $i$ as
\begin{equation}
P(T_i|i) = \prod_{w\in T_i}\frac{\exp (\bm{w} \cdot \bm{i})}{\sum_{w'\in V}\exp (\bm{w}' \cdot \bm{i})}
\label{equ:item_embedding}
\end{equation}
where $\bm{w}\in\mathbb{R}^{\alpha}$ is the embedding of word $w$, and $V$ is the vocabulary of all possible words.
Note that it is important to learn $\bm{i}$ separately instead of representing it by averaging word embeddings because user purchases could be affected by information other than text~\cite{ai2017learning}.

%In contrast to product items, we do not know the words of each query before a user actually issues it.
%Therefore, the embedding of queries can only be computed on the fly.
%Previous studies~\cite{van2016learning,ai2017learning} have explored several methods to construct query embeddings from query words. 
%As far as we know, the state-of-the-art method is to compute query embeddings with a non-linear projection function $\phi$ defined as
%\begin{equation}
%\phi(\{w_q | w_q \in q\}) = \tanh(W \cdot \frac{\sum_{w_q \in q}\bm{w_q}}{|q|} + b)
%\label{equ:fs}
%\end{equation}
%where $\bm{w_q}$ is the embedding of a word $w_q$ in $q$, $|q|$ is the length of the query, and $W\in\mathbb{R}^{\alpha \times \alpha}$ and $b\in\mathbb{R}^{\alpha}$ are two parameters learned in the training process.

%HEM
\textbf{Hierarchical Embedding Model}. 
Similar to QEM, HEM~\cite{ai2017learning} also computes query embeddings with the encoding function $\phi$ and item embeddings with their associated text $T_i$. 
However, in contrast to QEM, HEM defines $M_{uq}$ in Eq.~(\ref{equ:P_iuq}) as  
\begin{equation}
\bm{M_{uq}} = \bm{q} + \bm{u}
\label{equ:HEM_Muq}
\end{equation}
where $\bm{u}$ is the embedding representation of the user $u$.
In this way, HEM considers both query intents and user preferences in the ranking of items for product search.
%As shown in Eq. (\ref{equ:QEM_goal}), the naive version of QEM does not consider user information in the modeling of purchases in search.
%In practice, however, user purchases show high correlations with people's personal taste and preferences~\cite{ai2017learning,guo2018multi}.
%To address this problem, Ai et al.~\cite{ai2017learning} proposes a Hierarchical Embedding Model (HEM) for personalized product search by incorporating user as an additional factor in the modeling of search purchases.
%Let $u$ be the user who submitted a query $q$, then the probability of whether $u$ would purchase an item $i$ in HEM is defined as 
%\begin{equation}
%P(i|u,q) = \frac{\exp (\bm{i} \cdot M_{uq})}{\sum_{i'\in I_q}\exp (\bm{i}' \cdot \bm{M_{uq}})}
%\label{equ:HEM_goal}
%\end{equation} 
%where $M_{uq}$ is a joint model of user-query pair $(u,q)$ defined as
%\begin{equation}
%M_{uq} = \bm{q} + \bm{u}
%\end{equation}

%In HEM, the embeddings of item $i$ and query $q$ are obtained in the similar way with QEM as shown in Eq. (\ref{equ:item_embedding})\&(\ref{equ:fs}), and the embedding of user $u$ is learned by optimizing the likelihood of observed user text $T_u$ given $u$ as
In HEM, the embedding of user $u$ is obtained by optimizing the likelihood of observed user text $T_u$ given $u$ as
\begin{equation}
P(T_u|u) = \prod_{w\in T_u}\frac{\exp (\bm{w} \cdot \bm{u})}{\sum_{w'\in V}\exp (\bm{w}' \cdot \bm{u})}
\label{equ:HEM_u}
\end{equation}
where $T_u$ could be any text written or associated to $u$, such as product reviews or the descriptions of items that the user has purchased.

\begin{figure*}
	\centering
	\includegraphics[width=5.5in]{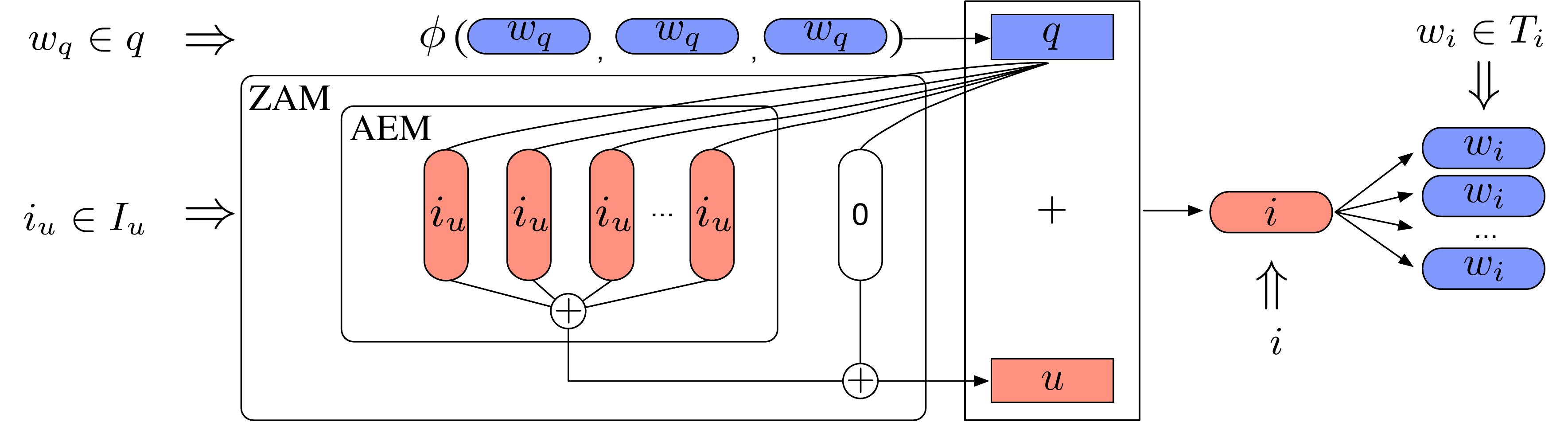}
	\caption{The structure of the Attention-based Embedding Model (AEM) and the Zero Attention Model (ZAM). $I_u$ is the purchase history of user $u$, $i$ is a candidate item for query $q$, and $w_q$ and $w_i$ are words in $q$ and the text associated with $i$ ($T_i$), respectively.}
	\vspace{-10pt}
	\label{fig:ZAM}
\end{figure*}

\subsection{Attention-based Personalization}\label{sec:attention_model}

%As discussed previously, QEM is an non-personalized product search model and HEM is an extension of QEM that incorporates use information for search personalization.
%As far as we know, HEM represents the state-of-the-art performance for personalized product retrieval models.
As discussed by Ai et al.~\cite{ai2017learning}, HEM is constructed based on the assumption that user preferences are independent of query intents in search.
This, however, is not true in practice.
For example, a customer who likes the baby products from \textit{Johnson's} may not want to buy \textit{Johnson's} baby shampoo when they search ``shampoo for men''.
To address this problem, a better paradigm for personalization in product search is to consider the relationships between queries and a user's purchase history.
%Previous studies~\cite{chen2018sequential,zhang2018towards} have shown that user profiling with attention mechanisms over purchased items can significantly improve the performance of product recommender systems.
Specifically, we apply an attention function over the user's purchase history to build user embeddings for product search.
Let $I_u$ be the set of items purchased by user $u$ before query $q$, then we can compute the embedding of $u$ as 
\begin{equation}
\bm{u} = \sum_{i \in I_u}\frac{\exp(f(q,i))}{\sum_{i' \in I_u}\exp(f(q,i'))}\bm{i}
\label{equ:AEM_u}
\end{equation}
where $f(q,i)$ is an attention function that determines the attention weight of each item $i$ with respect to the current query $q$.  
Similar to previous studies on attention models~\cite{cho2014learning,vinyals2015pointer}, we define $f(q,i)$ as
\begin{equation}
f(q, i) = \big(\bm{i}\cdot \tanh(\bm{W}_f \cdot \bm{q} + \bm{b}_{f})\big) \cdot \bm{W}_h
\label{equ:attention_function}
\end{equation}
where $\bm{W}_h\in \mathbb{R}^{\beta}$, $\bm{W}_{f} \in \mathbb{R}^{\alpha \times \beta \times \alpha}$, $\bm{b}_{f} \in \mathbb{R}^{\alpha \times \beta}$, and $\beta$ is a hyper-parameter that controls the number of hidden units in the attention network. 
Given the attention-based user embedding $\bm{u}$, we can further conduct personalized product search using the same method described in Eq.~(\ref{equ:P_iuq}) and Eq.~(\ref{equ:HEM_Muq}).
We refer to this model as the Attention-based Embedding Model (AEM). 

In contrast to HEM, AEM conducts personalization with query-dependent user profiling.
The embedding representations of each user are constructed according to their queries so that the model can better capture relevant user preferences in the current search context.
This is beneficial especially when the user has purchased many products that are irrelevant to the current query.
There is another attention-based product search model proposed by Guo et al.~\cite{guo2019attentive} that shares the similar idea of AEM.
Unfortunately, for the computation of attention weights, their model assumes that each item in a user's purchase history should associate with at least one search query submitted by users, which is not true in our datasets. Thus, we ignore it in this paper.

There is, however, an important problem that limits the power of both HEM and AEM.
As shown in Section~\ref{sec:log_analysis}, different queries have different potential for personalization.
Despite the efforts on query-dependent user profiling, the attention mechanism in Eq.~(\ref{equ:AEM_u}) requires AEM to attend to at least one item in the user's purchase history, which means that it always personalizes. 
Ai et al.~\cite{ai2017learning} have explored a naive solution that adds a hyper-parameter in Eq.~(\ref{equ:HEM_Muq}) to control the weight of user embedding $\bm{u}$ in $\bm{M_{uq}}$, but this merely trades off the gains of personalization on some queries with its loss on other queries.  
To actually solve this problem, we need a model that can automatically determine when and how to do personalization in product search.

% Problem of existing methods

% Idea of attention
% Model formulation

% Advantages and disadvantages
% query-dependent user modeling, fixed personalization weight

\subsection{Zero Attention Strategy}

%In this paper, we propose a Zero Attention Model (ZAM) for personalized product search.
%The goal of ZAM is to conduct differentiated personalization in product search, and it achieves so by applying a Zero Attention Strategy in the construction of user embeddings.
%In this section, we describe the details of ZAM as well as a theoretical analysis on how the Zero Attention Strategy controls the effect of search personalization.

The usefulness of personalization depends on both the query and the user's purchase history. 
For example, spear-fishing queries often result in the same item being purchased regardless of customer preferences. Similarly, a customer who is shopping in a certain category for the first time may have no relevant history on which to base personalization. To address this issue, we propose a Zero Attention Strategy that relaxes the constraints of existing attention mechanisms by introducing a Zero Vector in the attention process.
Accordingly, we propose a Zero Attention Model (ZAM) that conducts differentiated personalization in product search based on search queries and user's purchase history.
%AEM cannot handle these situations as its attention mechanism requires the model to pay attention on at least one item in the construction of user embeddings. 
%To address this problem, we propose to relax the constraints of existing attention mechanisms by introducing a Zero Vector in the attention process.

Figure~\ref{fig:ZAM} shows the structure of ZAM. 
Similar to AEM, ZAM learns item embeddings based on their associated words, and conducts retrieval with both query embeddings and user embeddings.
The main difference between ZAM and AEM is that, instead of attending to the user's previous purchases only, ZAM allows the attention network to attend to a Zero Vector, which we refer to as the Zero Attention Strategy.
Formally, let $\bm{0} \in \mathbb{R}^{\alpha}$ be the Zero Vector where each element is 0.
Then, in ZAM, the embedding representation of a user $u$ is computed as
\begin{equation}
\bm{u} = \sum_{i \in I_u}\frac{\exp(f(q,i))}{\exp(f(q, \bm{0})) + \sum_{i' \in I_u}\exp(f(q,i'))}\bm{i}
\label{equ:ZAM_u}
\end{equation}
where $f(q,\bm{0})$ is the attention score of $\bm{0}$ with respect to query $q$.
%We refer to this new attention mechanism as the Zero Attention Strategy (ZAS).

We now show how this simple modification achieves differentiated personalization in product search.
Let $\bm{x}\in \mathbb{R}^{|I_u|}$ be a vector formed from $\{f(q,i)|i\in I_u\}$. % for all items in $I_u$.
Then, Eq.~(\ref{equ:ZAM_u}) can be reformulated as
\begin{equation}
\bm{u} = \frac{\exp(\bm{x})}{\exp(f(q, \bm{0})) + \exp^+(\bm{x})}\cdot \bm{I_u}
\label{equ:ZAM_sig}
\end{equation}
where $\bm{I_u}$ is a matrix consisting of all item embeddings in the user's purchase history, and $\exp^+(\bm{x})$ is the element-wise sum of $\exp(\bm{x})$.
In Eq.~(\ref{equ:attention_function}), $\exp(f(q,\bm{0})) = 1$, so the factor of $\bm{I_u}$ in Eq.~(\ref{equ:ZAM_sig}) is actually a sigmoid function of $\bm{x}$.
In other words, the introduction of the Zero Attention Strategy creates an activation function that controls the influence of user purchase history in the current search context.
The value of $\exp^+(\bm{x})$ is the cumulative attention received by the user's previous purchased items given the query, and $\exp(f(q,\bm{0}))$ is essentially a threshold for personalization. Although our formulation is constant, this threshold can be query-dependent by defining $f$ with a more complicated function.
In any case, the user embedding $\bm{u}$ should only have strong influence in ZAM if the user shows consistent and salient interests on products related to the current query.
This enables ZAM to conduct differentiated personalization in different search scenarios.

\subsection{Model Optimization}

Similar to previous studies~\cite{van2016learning,ai2017learning,wu2018turning}, we optimize both AEM and ZAM by maximizing the log likelihood of observed user purchases and item information.
Specifically, we represent each item with their titles, each user with their previously purchased items, and each query with their query words.
Let $T_i$ be a list of words in the title of item $i$, then the log likelihood of an observed user-query-item triple can be computed as
\begin{equation}
\begin{split}
\mathcal{L}(T_i,u,i,q) =&\log P(T_i|i) \!+\! \log P(i|u,q) \!+\! \log P(u,q) \\
\approx& \sum_{w_i\in T_i}\log\frac{\exp (\bm{w}_i \cdot \bm{i})}{\sum_{w'\in V}\exp (\bm{w}' \cdot \bm{i})} \\
&+ \log \frac{\exp (\bm{i} \cdot (\bm{q} + \bm{u}))}{\sum_{i'\in I_q}\exp (\bm{i}' \cdot (\bm{q} + \bm{u}))}
\end{split}
\label{equ:log_likelihood}
\end{equation}
where $\bm{w}$ and $\bm{i}$ are learned parameters, $\bm{q}$ is computed with Eq.~(\ref{equ:fs}), $\bm{u}$ is computed with Eq.~(\ref{equ:AEM_u}) (for AEM) or Eq.~(\ref{equ:ZAM_u}) (for ZAM), and $\log P(u,q)$ is ignored because training examples are sampled uniformly from the logs.
%given a uniform distribution assumption on search sessions.
%Given a uniform assumption on the distribution of each search session in the training data, $\log P(u,q)$ can be ignored in the ranking optimization process. 

Computing $\mathcal{L}(T_i,u,i,q)$, however, is often infeasible because of the large number of words in $V$ and items in $I_q$.
For efficient training, we adopt a negative sampling strategy~\cite{mikolov2013efficient,le2014distributed,ai2016analysis} to estimate Eq.~(\ref{equ:log_likelihood}).
Specifically, for each softmax function, we sample $k$ negative samples to approximate its denominator.
The final optimization function of AEM and ZAM can be formulated as
\begin{equation}
\begin{split}
\mathcal{L}' =& \sum_{(u,q,i)}\mathcal{L}(T_i,u,i,q) \\
\approx&\!\!\!\!\sum_{(u,q,i)}\sum_{w_i \in T_i}\big(\log \sigma(\bm{w_i} \!\cdot\! \bm{i}) \!+ \!k\cdot \mathbb{E}_{w'\sim P_w}[\log\sigma(-\bm{w'} \!\cdot\! \bm{i})] \big) \\
&+\log \sigma\big(\bm{i} \cdot (\bm{q}+\bm{u})\big)\\ 
&+k\cdot \mathbb{E}_{i'\sim P_{I_q}}\big[\log\sigma\big(-\bm{i'} \cdot (\bm{q}+\bm{u})\big)\big]
\end{split}
\label{equ:final_loss}
\end{equation}
where $\sigma(x)$ is a sigmoid function, and $P_w$ and $P_{I_q}$ are the noise distribution of word $w$ and item $i$, respectively.
In our experiments, we define $P_w$ as the unigram distribution raised to the 3/4  power~\cite{mikolov2013efficient}, and $P_{I_q}$ as a uniform distribution.
%Also, as shown in Figure~\ref{fig:ZAM}, all the words in queries and item titles share a common embedding space, and all the items in $I_u$ and $I_q$ share another embedding space.
Also, as shown in Figure~\ref{fig:ZAM}, words in queries and item titles and items in $I_u$ and $I_q$ share a common embedding space.
We experimented with different types of regularization techniques such as L2 regularization, but none made a significant difference to the performance of the models. 
This indicates that, given the large-scale training data we have, overfitting is not an issue for model optimization in our experiments.
Thus, we ignore the regularization terms in Eq.~\ref{equ:final_loss} for simplicity.
%emprical explanations 
% now we show theory
% idea of zero attention

% model formulation

% Advantages and disadvantages
% Theoretical analysis
% THreshold theory

%!TEX root=CIKM19-ZAM.tex
\section{Experimental Setup}\label{sec:setup}

In this section, we describe our experimental settings for the log analysis and product search experiments.
We first introduce how we collect search logs from a commercial product search engine, and then the details of our baselines and the parameter settings for different product retrieval models.

\subsection{Datasets and Evaluation}

%product category
Our log analysis and retrieval experiments are carried on products from three categories -- \textit{Beauty}, \textit{Grocery}, and \textit{Health\&Personal Care}.
\textit{Beauty} consists of products for beauty purposes, such as facial cleanser and hair care; \textit{Grocery} consists of food purchases like chips and vegetables; and \textit{Health\&Personal Care} contains more items on non-beauty personal care, such as vitamins, dental cleaning products, etc.
Any search sessions result in a purchase on products that belong to a certain category would be classified as a search in that category. 
%in this paper we focus on personalization with previous purchases.

\begin{table}[t]
	\centering
	\caption{Some basic statistics of our experimental datasets.}
	\begin{tabular}{ c  | c | c | c     } %p{5mm}
		\hline
		&	Beauty  &	Grocery &	Hpc \\\hline
        Unique Words & 168,831 & 132,032 & 191,640\\
        Unique Users &	999,130	&998,655&	997,989\\
        Unique Queries	&1,369,893	&918,453	&1,171,631\\
        Unique Products	&3,940,175	&2,928,612	&3,868,178\\
        Unique Sessions	&3,985,150	&4,381,715	&3,809,994\\
%Test Session Number	500820	505187	462699

		\hline
	\end{tabular}
	\label{tab:data_stat}
\end{table}

%log sample strategy
%To construct our experimental test bed, we collected the search logs of a commercial e-commerce search engine from Jun. 5th to Jun. 10th in 2018.
%We randomly sampled 1 million customers for each product category to form our experiment user sets.
To construct our experimental testbed and preserve the privacy of customers at the same time, we anonymize all customer information beforehand and randomly sampled one million customers for each product category from the search logs of a commercial e-commerce search engine in 2018. %Jun. 5th to Jun. 10th 2018.
We collected the search sessions of each user in the corresponding category from a one year period. %Sep. 1st 2017 to Jun. 10th 2018.
We also collected the top 100 items retrieved by an initial product retrieval system for each query as our candidate item sets.
Each item is represented with its title and a unique product ID. 
In this paper, we focus on predicting user purchases based on the user's current query and their purchase histories. 
Only the purchased items are considered relevant in each session, and we manually exclude the sessions with no purchase.
Search sessions are divided with 30-minutes inactivity or query changing.
%Our ground truth is user purchases in search, where only the purchased items in a session are treated as relevant.
Our training and test sets are formed by splitting search sessions chronologically on a specific date to achieve a train/test ratio of 7 to 1. %from the first and the second week of the sampled log period, respectively. %Sep. 1st 2017 to Jun. 5th 2018 and Jun. 5th 2018 to Jun. 10th 2018, respectively.
The history of a user is formed by their purchases and queries in the corresponding category before the current search session.  
Test users that have never appeared in the training set were removed. 
We also remove words that have appeared less than 5 times in the whole dataset and simply ignore OOV words in the training and testing process.
%but all queries in the test set were retained, including those that did not appear in the training set.
The basic statistics of our datasets are shown in Table~\ref{tab:data_stat}.
%However, in order to evaluate the ability of the model to generalize to new queries, all queries in the test set were retained, including those that did not appear in the training set.
%In total, we have more than 300K sessions in the test set of each product category.

For ranking evaluation, we compute Mean Reciprocal Rank (MRR), Normalized Discounted Cumulative Gain at 10 (NDCG@10), and Hit ratio at 10 (Hit@10) in session level.
MRR focuses on the position of the first retrieved item that the user has purchased; NDCG evaluates the overall ranking performance over all purchased items;  %(there are many sessions where users purchased more than one items); 
%MRR and NDCG are two standard ranking metrics widely used in information retrieval (IR) studies, 
and Hit@10 is a recall-focused metric that measures the percentage of sessions where users have purchased at least one item in the top 10 results.
Note that 
Significant differences are computed based on student t-test with $p\leq 0.01$, and model performances are reported as improvement percentages over the standard language modeling approach for IR (i.e. the Query Likelihood model).

% data split

% evaluation

\subsection{Baselines}

In our experiments, we include six types of product search baselines:
%classic term-based retrieval model, statistic popularity models, and embedding-based neural retrieval models:
\begin{itemize}
\item \textbf{Query Likelihood Model} (QL): The standard language modeling approach for IR~\cite{ponte1998language}, which ranks products based on the log likelihood of queries in the item's unigram language model constructed with title words.
\item \textbf{Query-dependent Popularity} (Pop$_q$): The popularity model that ranks items according to how many times they have been purchased given the query $q$ in the training data. 
\item \textbf{Query-independent Popularity} (Pop$_i$): The popularity model that ranks items according to how many times they have been purchased in all sessions in the training data. 
\item \textbf{Query Embedding Model} (QEM): The product retrieval model based on the embedding-based generative framework described in Section~\ref{sec:framework}, which can be treated as a non-personalized version of the Hierarchical Embedding Model.
\item \textbf{Hierarchical Embedding Model} (HEM): The personalized product search model proposed by Ai et al.~\cite{ai2017learning} described in Section~\ref{sec:framework}. %HEM constructs user profiles from the text associated with each user, such as the titles, descriptions, or reviews of purchased items. 
In this paper, we use the purchased item's titles as the text associated with each user.
\item \textbf{Attention-based Embedding Model} (AEM): A personalized product search model proposed in Section~\ref{sec:attention_model}, which constructs query-dependent user embeddings with a classic attention mechanism over the user's purchase history.
\end{itemize}
Among the six baseline methods above, QL is a classic term-based retrieval model, Pop$_q$/Pop$_i$ are statistical models, and QEM, HEM and AEM are embedding-based neural retrieval models.
The first four are non-personalized product search models, while the last two are personalized product search models.
%As shown in Section~\ref{sec:model}, the only difference between the personalized product search baselines (i.e., HEM and AEM) and our proposed Zero Attention Model (ZAM) is 

% consider other information is out of the scope of this paper

\subsection{Parameter Settings}

Both the neural baselines and our Zero Attention Model (ZAM) are trained on an NVIDIA Tesla K40 GPU with 12 GB memory. 
We initialized all embedding vectors randomly, and trained each neural model for 20 epochs in total (most models converged after 10 epochs).
We used the Adagrad optimizer with an initial learning rate of 0.5.
We set the batch size as 256 and the number of negative samples $k$ as 5.
For QL, we tuned the Dirichlet smoothing factor $\mu$ from 10 to 100 (the average length of item titles is shorter than 100).
For AEM and ZAM, we tuned the number of hidden units $\beta$ (in Section~\ref{sec:attention_model}) from 1 to 5, and set it as 3 in the final results.
And for all embedding models, we tuned the embedding size $\alpha$ from 100 to 300.
Large embedding size did not introduce any significant performance improvements in our experiments, so we only report the results with $\alpha=100$ for all embedding models.
Note that all model hyper-parameters are tuned on the training data.
We will release our code after the paper is published.
%The code will be released 
%computation environment
%QL mu, 
%embedding size, batch size, learning rate, optimizer, 

%!TEX root=CIKM19-ZAM.tex

\newcommand{\ExampleQuery}{vitamin c serum}
\newcommand{\QHPurchase}{claire vitamin c serum ... eye treatment  1 fl  oz}
\newcommand{\HAPurchase}{tolb vitamin c serum for face ... skin  1 fl oz}
\newcommand{\AZPurchase}{tolb vitamin c serum for face ... skin  1 fl oz}
\newcommand{\HAHistoryPurchase}{super vitamin c serum for face ... huge 2 fl oz}
\newcommand{\AZHistoryPurchase}{baebodi eye gel for dark circl ...  1.7 fl oz}

\begin{table*}[t]
	\centering
	\small
	\caption{The ranking performance of different product search models in each category. Metrics are reported as the improvement percentages over the best non-personalized baseline (i.e., QEM). $+$ and $-$ denote significant improvements or degradation over QEM, and $\ddagger$ denotes significant improvements over all the baselines. The best performance is highlighted in boldface.}
	\begin{tabular}{ c| c || c | c | c || c | c | c || c | c | c    } %p{5mm}
		\hline
		\multicolumn{2}{c||}{ } & \multicolumn{3}{c||}{\textit{Beauty}} & \multicolumn{3}{c||}{\textit{Grocery}} & \multicolumn{3}{c}{\textit{Health \& Personal Care}}\\ \hline 
		%\multicolumn{1}{c||}{ } & \multicolumn{3}{c||}{Dataset 1} & \multicolumn{3}{c}{Dataset 2}\\ \hline \hline
		\multicolumn{2}{c||}{Model}  & MRR & NDCG@10 & Hit@10 & MRR & NDCG@10 & Hit@10 & MRR & NDCG@10 & Hit@10  \\\hline
		\hline
		\multirow{4}{*}{Non-personalized} & QL & -62.59\%$^-$ & -63.19\%$^-$ & -56.17\%$^-$ & -70.86\%$^-$ & -70.76\%$^-$ & -60.92\%$^-$ & -67.66\%$^-$ & -68.06\%$^-$ & -60.18\%$^-$ \\ \cline{2-11}
		& Pop$_q$  & -49.34\%$^-$ & -47.32\%$^-$ & -35.81\%$^-$ & -50.61\%$^-$ & -46.84\%$^-$ & -32.10\%$^-$ & -55.16\%$^-$ & -52.36\%$^-$ & -38.68\%$^-$ \\ \cline{2-11}
		& Pop$_i$ & -27.08\%$^-$ & -24.70\%$^-$ & -16.05\%$^-$ & -31.78\%$^-$ & -27.36\%$^-$ & -14.66\%$^-$ & -34.13\%$^-$ & -29.28\%$^-$ & -15.45\%$^-$ \\ \cline{2-11}
		& QEM & 0\% & 0\% & 0\% & 0\% & 0\% & 0\% & 0\% & 0\% & 0\%\\ \hline \hline
		\multirow{3}{*}{Personalized} & HEM & -0.94\%$^-$ & -1.38\%$^-$ & -1.79\%$^-$ & 6.05\%$^+$ & 4.70\%$^+$ & 1.48\%$^+$ & -0.43\% & -1.18\%$^-$ & -2.23\%$^-$ \\ \cline{2-11}
		& AEM & 0.96\%$^+$ & 0.51\% & -0.29\% & 8.50\%$^+$ & 6.80\%$^+$ & 2.54\%$^+$ & 2.71\%$^+$ & 1.85\%$^+$ & 0.07\%\\ \cline{2-11}
		& \textbf{ZAM} & \textbf{2.77\%}$^{+\ddagger}$ & \textbf{2.10\%}$^{+\ddagger}$ & \textbf{0.59\%} & \textbf{9.46\%}$^{+\ddagger}$ & \textbf{7.60\%}$^{+\ddagger}$ & \textbf{2.91\%}$^{+}$ & \textbf{4.57\%}$^{+\ddagger}$ & \textbf{3.36\%}$^{+\ddagger}$ & \textbf{0.71\%}$^{+\ddagger}$\\ 
		\hline
	\end{tabular}
	\label{tab:overall_results}
	\vspace{-10pt}
\end{table*}

\section{Results and Discussion}\label{sec:results}

We now report our experimental results on personalized product search.
We first describe the overall performance of different non-personalized baselines and personalized product retrieval models, and then discuss the effect of personalization on different query groups. 
After that, we conduct a case study to shed some light on the advantages and disadvantages of each personalization method.

\subsection{Overall Performance}

Table~\ref{tab:overall_results} shows the overall performance of each product search model in our experiments.
As we can see, QL is significantly worse than other baselines in all categories, which is consistent with previous studies~\cite{van2016learning,ai2017learning}.
There are significant language gaps between how users formulate queries and how sellers write product descriptions~\cite{nurmi2008product}. In addition, simple queries will often map to a large number of items (e.g, ``phone case''), which makes it difficult for a term-based retrieval model to differentiate the relevance of these products.
%Additional evidence is that Pop$_q$, a simple baseline that ranks products by popularity, beats QL with approximately 50\% improvement in Table~\ref{tab:overall_results}.
In contrast, QEM achieves the best performance among the four non-personalized product retrieval models and is nearly two times better than QL.
This demonstrates the effectiveness of the embedding-based generative models in product retrieval.

%baselines
%QL vs others
%QEM vs others

If we compare the performance of personalized product search models with QEM, we can see that HEM is better than QEM on \textit{Grocery}, but significantly worse on \textit{Beauty} and \textit{Health \& Personal Care}.
As discussed in Section~\ref{sec:log_analysis}, different queries have different potential for personalization.
HEM conducts personalization by adding query-independent user embeddings to every query equally, which can hurt the system performance when a user's global purchase preferences are not related to their current search.
After applying the attention mechanism, AEM conducts query-dependent personalization and achieves better ranking performance than HEM.
Nonetheless, it still personalizes all queries with equal weights and has a limited improvement over QEM in \textit{Beauty}. %on queries where personalization is not important.
%We will further discuss this phenomenon in the next section.

In our experiments, ZAM performs the best among all product search models. 
On all categories, ZAM achieves more than 2\% improvement over QEM on both MRR and NDCG@10.
It also significantly outperforms HEM and AEM.
This demonstrates that the introduction of the Zero Attention Strategy indeed improves the overall effectiveness of personalized product search models.

%HEM vs QEM

%AEM and ZAM vs QEM and HEM
\begin{figure*}
	\centering
	\begin{subfigure}{.33\textwidth}
		\centering
		\includegraphics[width=2.2in]{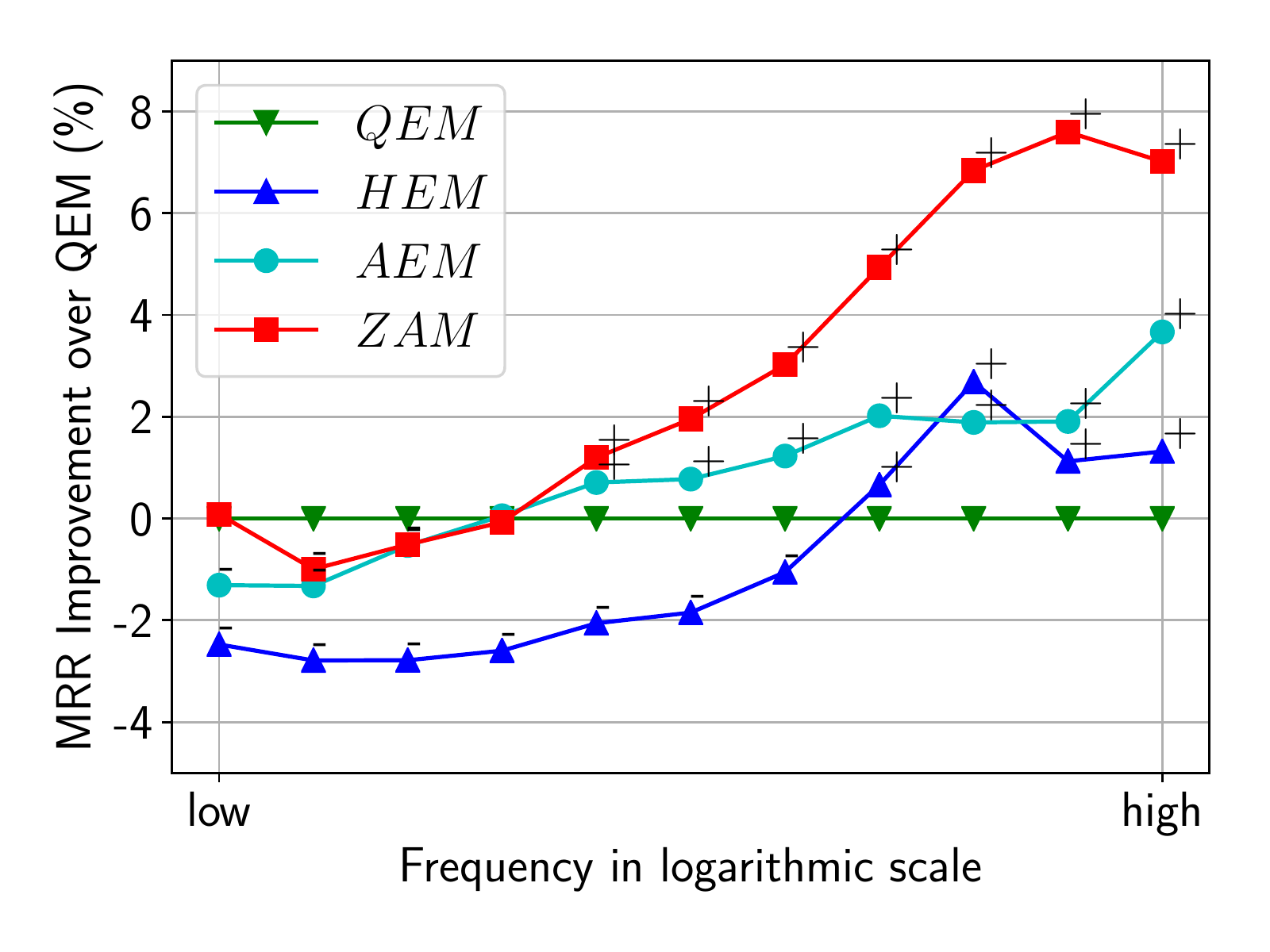}
		\vspace{-10pt}
		\caption{\textit{Beauty}}
		\label{fig:b_frequency}
	\end{subfigure}%
	\begin{subfigure}{.33\textwidth}
		\centering
		\includegraphics[width=2.2in]{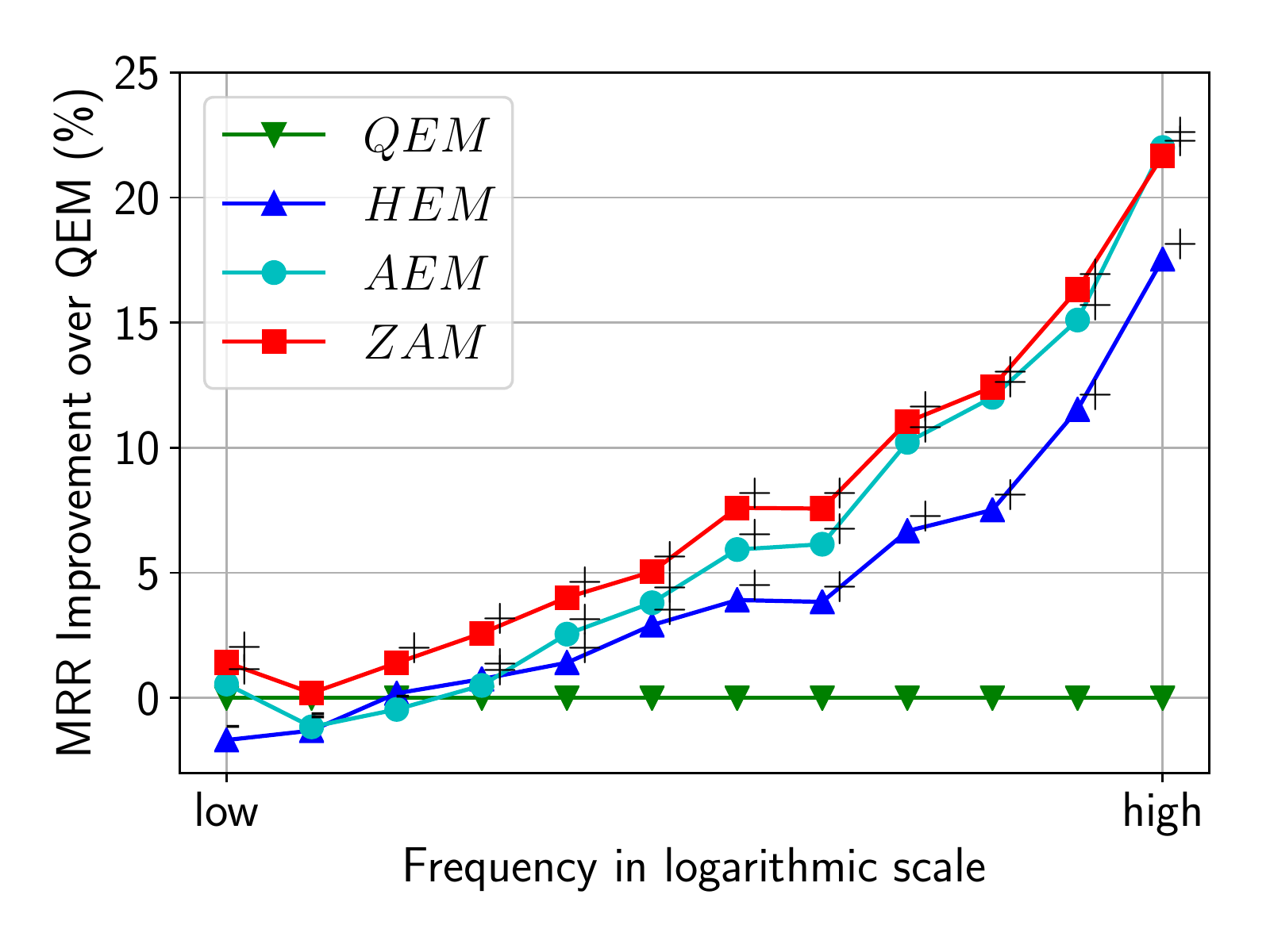}
		\vspace{-10pt}
		\caption{\textit{Grocery}}
		\label{fig:g_frequency}
	\end{subfigure}%
	\begin{subfigure}{.33\textwidth}
		\centering
		\includegraphics[width=2.2in]{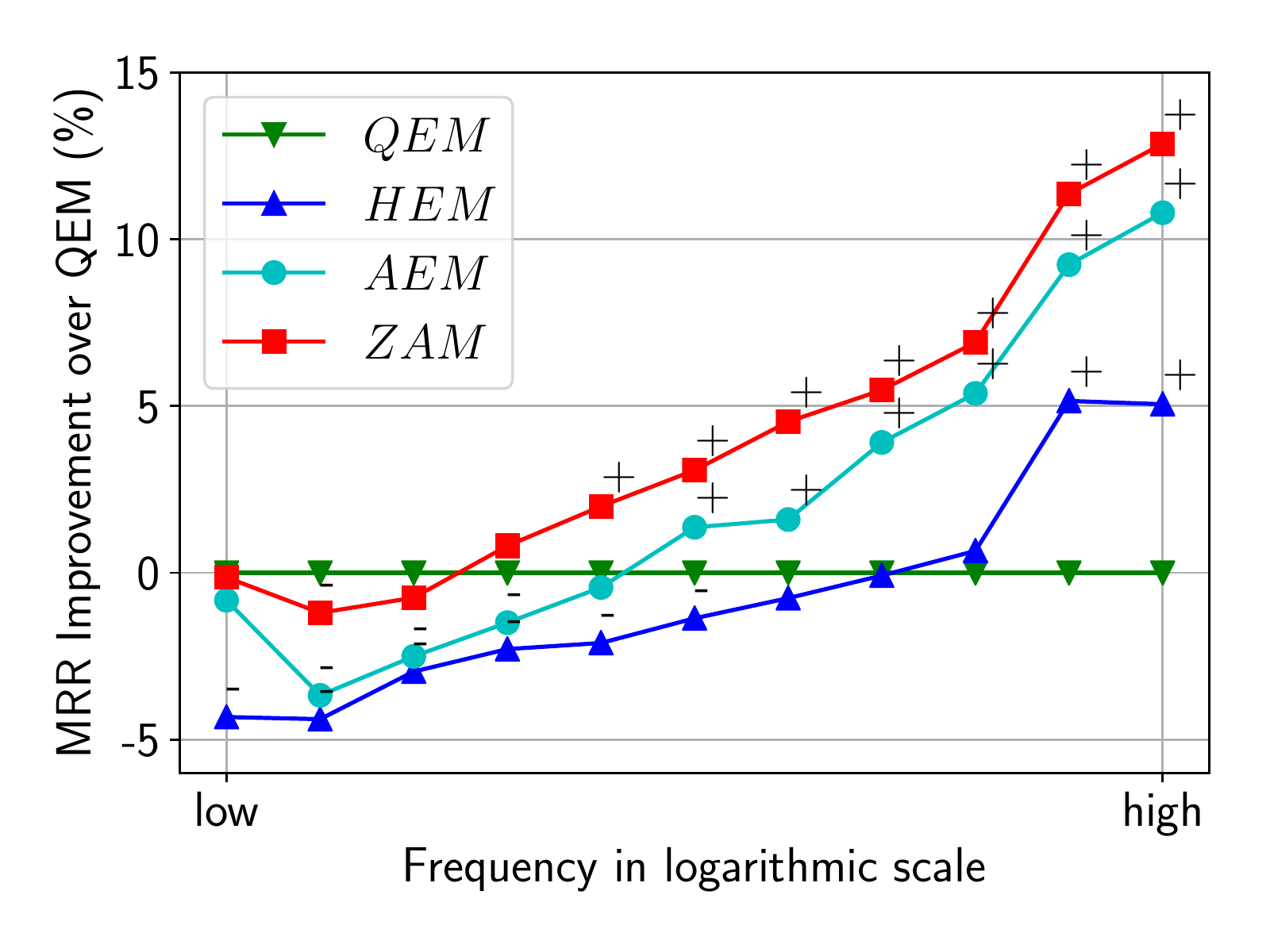}
		\vspace{-10pt}
		\caption{\textit{Health\&Personal Care}}
		\label{fig:h_frequency}
	\end{subfigure}%
	\vspace{2pt}
	\caption{The performance of different embedding-based product retrieval models with respect to query frequencies. $+$ and $-$ denote significant improvements or degradation with respect to QEM.}
	\vspace{-10pt}
	\label{fig:frequency}
\end{figure*}

\begin{figure*}
	\centering
	\begin{subfigure}{.33\textwidth}
		\centering
		\includegraphics[width=2.2in]{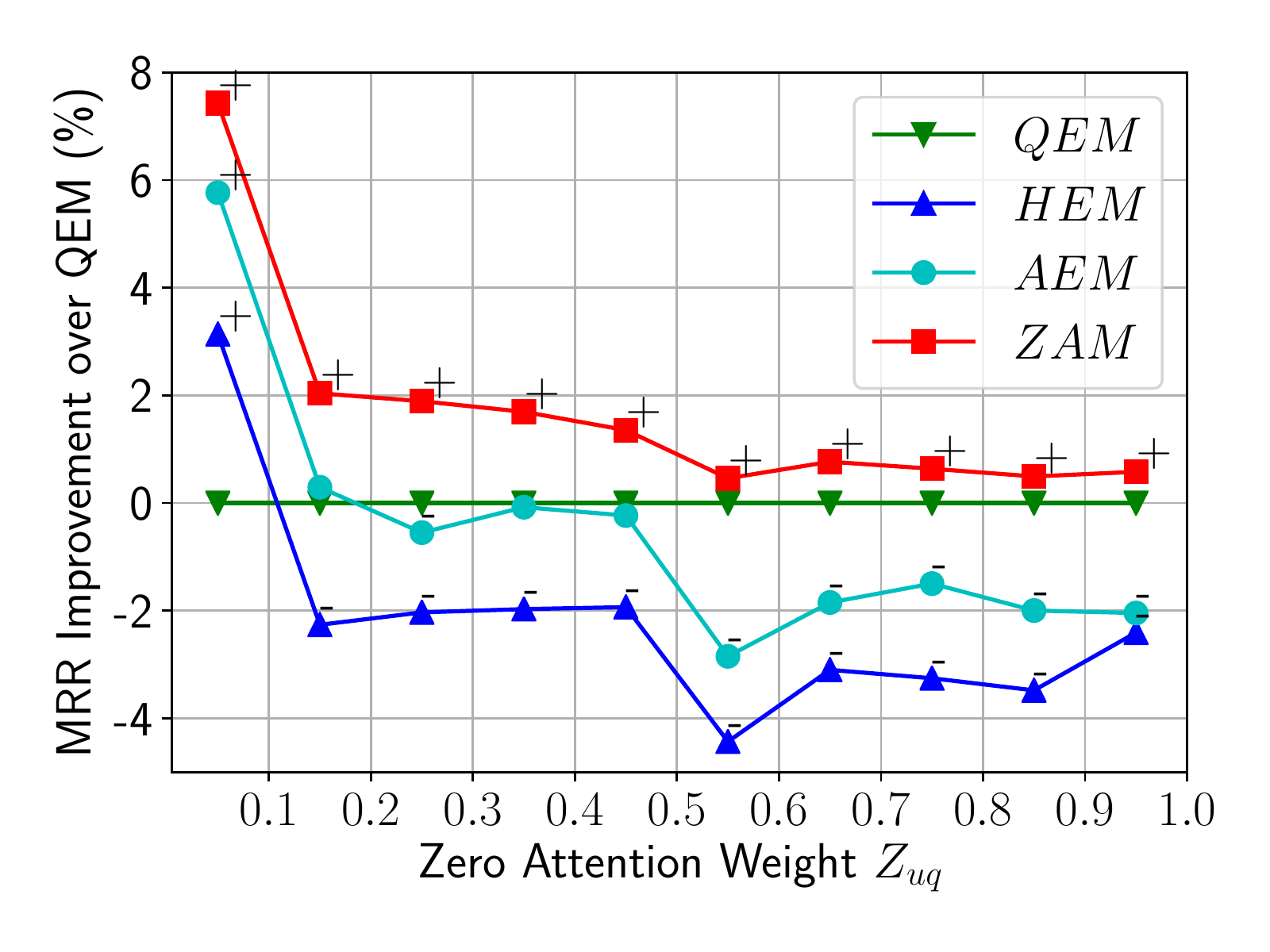}
		\vspace{-10pt}
		\caption{\textit{Beauty}}
		\label{fig:b_att}
	\end{subfigure}%
	\begin{subfigure}{.33\textwidth}
		\centering
		\includegraphics[width=2.2in]{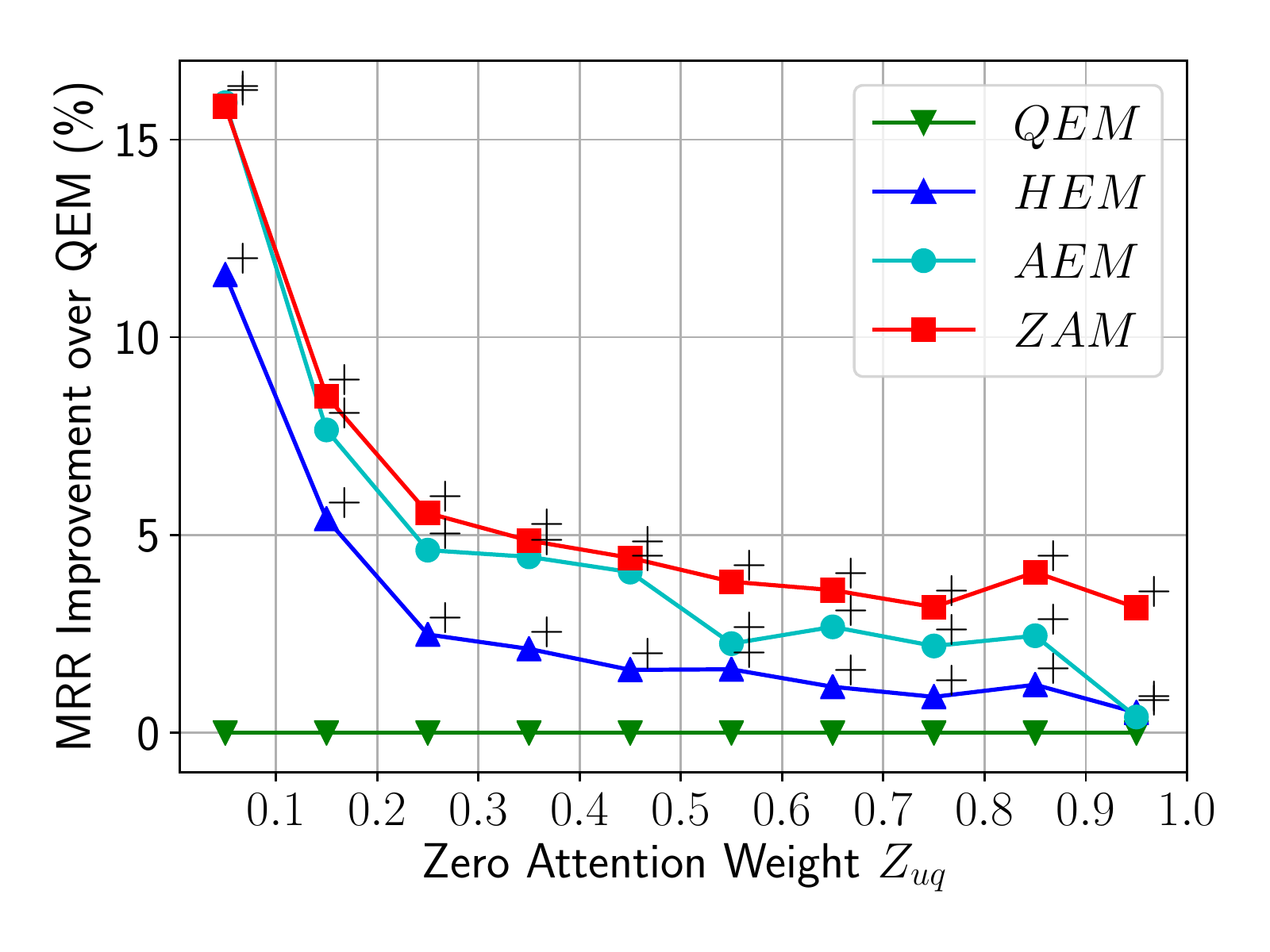}
		\vspace{-10pt}
		\caption{\textit{Grocery}}
		\label{fig:g_att}
	\end{subfigure}%
	\begin{subfigure}{.33\textwidth}
		\centering
		\includegraphics[width=2.2in]{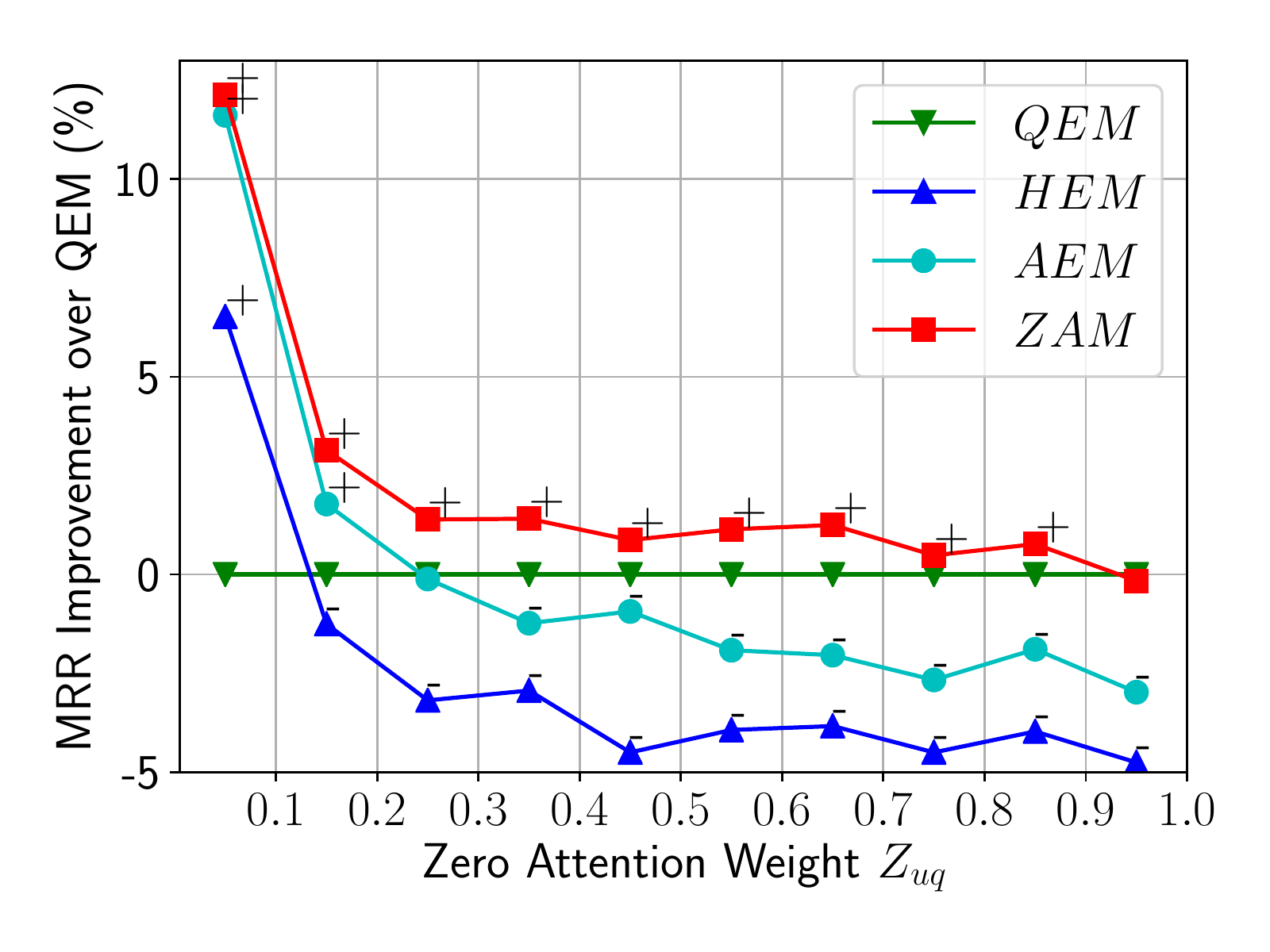}
		\vspace{-10pt}
		\caption{\textit{Health\&Personal Care}}
		\label{fig:h_att}
	\end{subfigure}%
	\vspace{2pt}
	\caption{The performance of different embedding-based product retrieval models with respect to the Zero Attention Weight $Z_{uq}$ in Eq.~(\ref{equ:ZeroAttentionWeight}). $+$ and $-$ denote significant improvements or degradation with respect to QEM.}
	\vspace{-10pt}
	\label{fig:attWeight}
\end{figure*}

\subsection{Effect of Personalization}

In this section, we describe a quantitative analysis of how personalization changes the performance of product retrieval models on different query groups.
Figure~\ref{fig:frequency} depicts the MRR improvements of HEM, AEM, and ZAM with respect to QEM on queries with different frequencies.
In our experiments, QEM performs well and achieves the best performance on queries with low frequencies in most cases.
HEM and AEM, on the other hand, performs worse than QEM on these queries after the incorporation of user information.
As discussed in Section~\ref{sec:log_analysis}, low-frequency queries usually have a spear-fishing intent.
People who submit these queries are likely to have exact items in mind. 
Thus, in these cases, the risk of personalization is larger than its potential.
On queries with higher frequencies, we observe more sessions where personalized product search models outperform the non-personalized model (i.e., QEM), especially in \textit{Grocery}.
One explanation is that people often have strong personal tastes on products retrieved for the head queries of \textit{Grocery}. 
For example, while different customers may buy different flavors of snack chips when they searched for ``chips'', we observed that many of them would purchase the same chips that they have purchased previously.
Among different personalization models, ZAM always produces the best performance and, in most cases, is significantly better than HEM and AEM.

%QEM is good on low
%ZAM is consistently better

In ZAM, we naturally create an activation function that controls the effect of personalization by allowing the model to attend to the Zero Vector $\bm{0}$. 
The attention weights of $\bm{0}$ in Eq.~(\ref{equ:ZAM_u}), namely the Zero Attention Weight, could be a valuable indicator for the potentials of personalization in each product search query.
We define the Zero Attention Weight of a user-query pair $(u,q)$ in ZAM as
\begin{equation}
\begin{split}
%Z_{uq} &= 1- \sum_{i \in I_u}\frac{\exp(f(q,i))}{\exp(f(q, \bm{0})) + \sum_{i' \in I_u}\exp(f(q,i'))} \\
%&= \frac{\exp(f(q,\bm{0}))}{\exp(f(q, \bm{0})) + \sum_{i' \in I_u}\exp(f(q,i'))}.
%Z_{uq} &= \frac{\exp(f(q,\bm{0}))}{\exp(f(q, \bm{0})) + \sum_{i' \in I_u}\exp(f(q,i'))}.
Z_{uq} &= 1- \sum_{i \in I_u}\frac{\exp(f(q,i))}{\exp(f(q, \bm{0})) + \sum_{i' \in I_u}\exp(f(q,i'))}
\end{split}
\label{equ:ZeroAttentionWeight}
\end{equation}
When $Z_{uq}$ is 0, ZAM conducts maximum personalization on the query, which means that the query has great potentials for personalization; when $Z_{uq}$ is 1, ZAM conducts no personalization. 
To verify this hypothesis, we plot the performance of both non-personalized and personalized embedding models with respect to $Z_{uq}$ in Figure~\ref{fig:attWeight}.

\begin{table*}[t]
	\centering
	\small
	\caption{Example sessions on the query of ``\ExampleQuery'' and the recent purchases of each user (sorted by time) in \textit{Beauty}.}
	\scalebox{0.92}{
		\begin{tabular}{ p{1.5in}| c | c || p{1.5in}| c | c || p{1.5in}| c | c     } %p{5mm}
			\hline
			%\multicolumn{2}{c||}{ } & \multicolumn{3}{c||}{\textit{Beauty}} & \multicolumn{3}{c||}{\textit{Grocery}} & \multicolumn{3}{c}{\textit{Health \& Personal Care}}\\ \hline 
			%\multicolumn{1}{c||}{ } & \multicolumn{3}{c||}{Dataset 1} & \multicolumn{3}{c}{Dataset 2}\\ \hline \hline
			\multicolumn{3}{c||}{Session 1: ``\ExampleQuery''} & \multicolumn{3}{c||}{Session 2: ``\ExampleQuery''} & \multicolumn{3}{c}{Session 3: ``\ExampleQuery''}  \\\hline
			\hline
			
			%Purchased Item & \multicolumn{2}{c||}{Position in list} \\
			& QEM & HEM & & HEM & AEM & & AEM & ZAM \\ \hline \hline
			Purchased Item & \multicolumn{2}{c||}{Rank in list} & Purchased Item & \multicolumn{2}{c||}{Rank in list} & Purchased Item & \multicolumn{2}{c}{Rank in list} \\ \hline
			\textbf{\QHPurchase} & 7 & 1 &  \textbf{\HAPurchase}
			& 19 & 2 & \textbf{\AZPurchase}
			& 7 & 2\\ \hline \hline
			
			Previous Purchases of the User & \multicolumn{2}{c||}{Attention} & Previous Purchases of the User & \multicolumn{2}{c||}{Attention} & Previous Purchases of the User & \multicolumn{2}{c}{Attention} \\ \hline
			6 plastic afro pik ... lift hair comb detangle
			& $-$ & $-$ & perfume bottle travel refillable ... perfume atomizer spray & $-$ & $<$0.001 & \textbf{\AZHistoryPurchase} & \textbf{0.744}
			& 0.0325
			\\ \hline
			essie nail polish, cuticle car... primers and finishers & $-$ & $-$ &  \textbf{\HAHistoryPurchase} & $-$ & \textbf{0.964} & makeup brush set ... 32 pieces professional makeup & $<$0.001 & $<$0.001 \\ \hline
			kiss products ... piece nail kit, just for you & $-$ & $-$
			& labs under eye serum,drk circl, 1 FZ & $-$ & $<$0.001 & roller for face and body ... microneedle kit & 0.198 & $<$0.001 
			\\ \hline
			%no splash nail clippers ... for fingernail and toenail & $-$ & $-$ & batiste dry shampoo, blush fragrance, 3 Count & $-$ & $<$0.001 & exfoliating dry or wet ... body wash brush set for dry brushing & $<$0.001 & $<$0.001\\\hline
			%5 pcs nail tip glue ... adhesive super bond for acrylic nails tips & $-$ & $-$ &  crystal mineral deodorant spray ... white tea 4.0 oz & $-$ & $<$0.001 & applicator self tanning mitts with snug ... and rxfoliator glove & $<$0.001 & $<$0.001\\ \hline
			%botanical beauty ... therapeutic grade organic peppermint & $-$ & $-$ & rose water ... therapeutic grade bulgarian & $-$ & $<$0.001 & liquid flawless ... darker luxurious & 0.050 & $<$0.001\\ \hline
			\textbf{\QHPurchase} & $-$ & $-$
			&   premium retinol cream face moisturizer with hyaluronic
			& $-$
			& 0.034 & \textbf{Zero Vector} $\bm{0}$ & $-$ & \textbf{0.960} \\ \hline
			\hline
		\end{tabular}
	}
	\label{tab:case_study}
\end{table*}

In Figure~\ref{fig:attWeight}, the y axis is the MRR improvement of each model with respect to QEM, and the x axis is the $Z_{uq}$ in ZAM ranged from 0 to 1.
Note that $Z_{uq}$ is not a hyper-parameter but a variable automatically developed by ZAM during  training.
As shown in Figure~\ref{fig:b_att} and \ref{fig:h_att}, QEM significantly outperforms HEM and AEM on queries where $Z_{uq}$ is larger than 0.5, while HEM and AEM achieve large improvements over QEM on queries with $Z_{uq}$ is smaller than 0.1, even when the overall performance of HEM is significantly worse than QEM. % on \textit{Beauty} and \textit{Health\&Personal Care}.
Also, in Figure~\ref{fig:g_att}, we observe that the relative improvements of HEM and AEM with respect to QEM show a strong negative correlation with $Z_{uq}$.
This demonstrates that the weight on the Zero Vector $\bm{0}$ in ZAM is indeed a good indicator of whether a product search query is suitable for personalization.

\subsection{Case Study}

To understand how different personalization methods benefit or damage the performance of product retrieval, we conduct case studies on \textit{Beauty} and extract three representative sessions for ``\ExampleQuery" in Table~\ref{tab:case_study}.
For each session, we show the final purchased item, its position in the ranked list of each model, and the most recent search purchases of the user. % in \textit{Beauty}. %(from new to old).
%We also show the ranks of the purchased items in the ranked lists produced by each model. %, and the lower the rank is, the better the model performs. % in terms of predicting user purchases in product search.
In Session 2 and 3, we also show the attention weights on each previously purchased item computed by AEM and ZAM.
For privacy concerns, we randomly replace the item titles in Table~\ref{tab:case_study} with titles from other products with similar properties so that no user information would be revealed.
%Item information is anonymized by randomly replacing their titles with other items from similar categories so that 

Session 1 in Table~\ref{tab:case_study} is a typical example where personalized models (i.e., HEM) outperform non-personalized models (i.e., QEM).
As shown in the table, before the start of Session 1, the user had already purchased ``\QHPurchase'' in a former search session. 
Without personalization, QEM cannot use this information, and it ranks the item at position 7 for this user.
HEM, on the other hand, identifies that ``\QHPurchase'' is relevant to the user and it is likely that the user would purchase it again. %when it appears on the result page.
Thus, HEM ranks it at position 1 in Session 1.
This example indicates that personalization is particularly beneficial for re-finding or repeated purchase intents.
%This example indicates that, for re-finding or repeated purchase intents, personalization is beneficial as it can help the system to quickly identify the exact item that the user wants. 
%This means that personalization is beneficial for re-finding or repeated purchase intents. 

Session 2 in Table~\ref{tab:case_study} shows an example where the introduction of attention mechanisms improves the performance of personalization.
In HEM, all purchased items in a user's purchase history have equal importance for the construction of the user profile.
In Session 2, however, most of the user's previous purchases are not directly related to the current query.
As a result, HEM fails to promote the rank of the final purchased item ``\HAPurchase'' and places it at position 19.
In contrast, AEM builds user embeddings based on the relevance of each item with respect to the current query.
It identifies that ``\HAHistoryPurchase'' is the only item in the user's purchase history that is relevant to ``\ExampleQuery'', and thus allocates 96.4\% of the attention weights to the item.
Since ``\HAHistoryPurchase'' was no longer available on our experimental e-commerce website and ``\HAPurchase'' is a common alternative to it, AEM puts ``\HAPurchase'' at the second position in the final rank list. 
%Since ``\HAPurchase'' is a popular alternative to ``\HAHistoryPurchase'' and the latter is no longer sold on our experimental e-commerce websites, AEM puts ``\HAPurchase'' at the second position in the final rank list. 
 
The last example in Table~\ref{tab:case_study}, Session 3, depicts how the Zero Attention Strategy benefits personalized product search.
In this session, none of the user's previous purchases are directly related to \ExampleQuery, which means that personalization is not likely to be helpful here.
Nonetheless, because the attention mechanism in AEM requires it to construct the user embedding with at least one item in the purchase history, AEM allocates considerable attention on the most ``relevant'' item -- ``\AZHistoryPurchase''. 
We expect a user's tastes in eye gel provide little information on their preferences for ``\ExampleQuery'', so AEM fails to achieve a good performance in Session 3.
With the Zero Attention Strategy, ZAM successfully recognizes that Session 3 is not suitable for personalization and places 96.0\% of the attention weights on the Zero Vector $\bm{0}$.
This limits the effect of personalization to only 4.0\% and results in the promotion of ``\AZPurchase'' from position 7 in AEM to position 2 in ZAM.

%\input{discussion}

%!TEX root=CIKM19-ZAM.tex
\section{Conclusion and Future Work}\label{sec:conclusion}

In this paper, we propose a Zero Attention Model (ZAM) for personalized product search.
The proposed model constructs user embeddings with a Zero Attention Strategy that enables it to conduct differentiated personalization for product search based on both the query characteristics and user purchase histories.
As shown in the experiments, the introduction of the Zero Attention Strategy not only improves the performance of personalized product search models, but also provides important information on the potential of personalization in each product search query.

%As a next step, we plan to evaluate the ablity of ZAM in incorporating more heterogeneous information for product search.
%Especially, we are interested in studying the potential of ZAM for explainable information retrieval.
As a next step, we are interested in studying the potential of ZAM for explainable information retrieval.
Although the threshold $f(q,\bm{0})$ in the activation function of ZAM (i.e., Eq.~(\ref{equ:ZAM_sig})) could be query-dependent, the attention function used in this paper (i.e., Eq.~(\ref{equ:attention_function})) actually creates the same values for $f(q,\bm{0})$ in all queries. 
%This design naturally produces comparable attention scores on each item in the purchase history, and this is essential for the generation of search explanations and the development of an explainable product retrieval model.
On the one hand, this design naturally makes the attention scores of each item comparable among different queries and sessions, which could be useful for the generation of search explanations and the development of an explainable product retrieval model.
On the other hand, learning query-dependent personalization threshold could be benefitial for the retrieval performance of the system. 
Besides, in our preliminary log analysis, we only analyze the potential of personalization with respect to query frequency. 
More analysis on other query characteristics such as query length and average candidate price would be interesting. 
For example, we observed that many customers have consistent price preferences when purchasing products in the \textit{Clothing} category, which could be useful for personalizing their search experience. 
%We will further explore those directions in a future study.  

%zam
%dicover

%future work
% in this paper, although f(q,0) could be query-dependent, we actually choose a attention function that have same value for f(q,0) for all queries. This is useful because it makes the attention scores comparable among different queries, which could be very useful for explanable product search

\iftrue
\section{Acknowledgments}
This work was supported in part by the Center for Intelligent Information Retrieval and in part by a gift from Amazon Search. Any opinions, findings and conclusions or recommendations expressed in this material are those of the authors and do not necessarily reflect those of the sponsor.
\fi

%\nocite{*}
\balance
\bibliographystyle{ACM-Reference-Format}
\bibliography{sigproc} 

\end{document}